\documentclass[%
 reprint,
 amsmath,amssymb,
 aps,
prb,
 superscriptaddress
]{revtex4-1}

\usepackage{graphicx}
\usepackage{dcolumn}
\usepackage{bm}
\usepackage{braket}
\usepackage{color}

\newcounter{num}
\newcommand{\Rnum}[1]{\setcounter{num}{#1}\Roman{num}}

\begin{document}

\preprint{APS/123-QED}

\title{Non-Bloch Band Theory of Non-Hermitian Systems}

\author{Kazuki Yokomizo}
\affiliation{Department of Physics, Tokyo Institute of Technology, 2-12-1 Ookayama, Meguro-ku, Tokyo, 152-8551, Japan}
\author{Shuichi Murakami}
\affiliation{Department of Physics, Tokyo Institute of Technology, 2-12-1 Ookayama, Meguro-ku, Tokyo, 152-8551, Japan}
\affiliation{TIES, Tokyo Institute of Technology, 2-12-1 Ookayama, Meguro-ku, Tokyo, 152-8551, Japan}%




%
\begin{abstract}
In spatially periodic Hermitian systems, such as electronic systems in crystals, the band structure is described by the  band theory in terms of the Bloch wave functions, which reproduce energy levels for large systems with open boundaries. In this paper, we establish a generalized Bloch band theory in one-dimensional spatially periodic tight-binding models. We show how to define the Brillouin zone in non-Hermitian systems. From this Brillouin zone, one can calculate continuum bands, which reproduce the band structure in an open chain.
As an example, we apply our theory to the non-Hermitian Su-Schrieffer-Heeger model. We also show the bulk-edge correspondence between the winding number and existence of the topological edge states.
%
\end{abstract}
\pacs{Valid PACS appear here}
\maketitle
%
%

The band theory in crystals is fundamental for describing electronic structure~\cite{Kittel1996}. By introducing the Bloch wave vector ${\bm k}$, the band structure calculated within a unit cell reproduces that of a large crystal with open boundaries. Here it is implicitly assumed that the electronic states are almost equivalent between a system with open boundaries and one with periodic boundaries, represented by the Bloch wave function with real ${\bm k}$. This is because the electronic states extend over the system.

Recently, non-Hermitian systems, which are described by non-Hermitian Hamiltonians have been attracting much attention. These systems have been both theoretically and experimentally studied in many fields of physics~\cite{Gonzalez2017,Kozii2017,Zyuzin2018,Shen2018v2,Molina2018,Yoshida2018,Carlstrom2018,Philip2018,Chen2018,Moors2019,Okugawa2019,Budich2019,Yang2019,Yoshida2019,Wu2019,San2016,Zeng2016,Li2018,Kawabata2018s,Guo2009,Ruter2010,Feng2011,Regensburger2012,Feng2013,Poli2015,Zhen2015,Zhao2015,Ding2015,Weimann2017,Hodaei2017,Chen2017,St2017,Bahari2017,Wang2018,Zhou2018,Parto2018,Zhao2018,Harari2018,Bandres2018,Pan2018,Jin2018,Malzard2018,Oztas2018,Kremer2019,Bliokh2019,Wang2019nat,Chen2019,Lee2014,Xu2017,Ashida2017,Gong2018,Nakagawa2018,Takata2018,Pan2019,Li2019,Liu2019,Runder2009,Zeuner2015,Mochizuki2016,Xiao2017,Hatano1996,Hatano1997,Hatano1998,Lourenco2018,Gong2018,Rosenthal2018,Luo2018,Wang2018v,Ezawa2019v1,Ezawa2019v2,Ezawa2019v3}. In particular, the bulk-edge correspondence has been intensively studied in topological systems. In contrast to Hermitian systems, it seems to be violated in some cases. The reasons for this violation have been under debate~\cite{Hu2011,Esaki2011,Lee2016,Leykam2017,Martinez2018,Ye2018,Shen2018,Yuce2018,Yin2018,Yuce2018v2,Kunst2018,Yao2018,Gong2018,Yao20182d,Kawabata2018b,Yuce2018v3,Kawabata2019,Jin2019,Wang2019,Borgnia2019,Ozcakmakli2019,Edvardsson2019,chLiu2019,Lee2019}.

One of the controversies is that in many previous works, the Bloch wave vector has been treated as real in non-Hermitian systems, similarly to Hermitian ones. In Ref.~\onlinecite{Yao2018}, it was proposed that in one-dimensional (1D) non-Hermitian systems, the wave number $k$ becomes complex. The value of $\beta\equiv{\rm e}^{ik}$ is confined on a loop on the complex plane, and this loop is a generalization of the Brillouin zone in Hermitian systems. In non-Hermitian systems, the wave functions in large systems with open boundaries do not necessarily extend over the bulk but are localized at either end of the chain, unlike those in Hermitian systems. This phenomenon is called the non-Hermitian skin effect~\cite{Yao2018}. Thus far, how to obtain the generalized Brillouin zone has been known only for simple systems. 

In this paper, we establish a generalized Bloch band theory in a 1D tight-binding model in order to determine the generalized Brillouin zone $C_\beta$ for $\beta\equiv{\rm e}^{ik}$, $k\in{\mathbb C}$. First of all, we introduce the ``Bloch'' Hamiltonian ${\cal H}\left(k\right)$ and rewrite it in terms of $\beta$ as ${\cal H}\left(\beta\right)$. Then the eigenvalue equation $\det\left[{\cal H}\left(\beta\right)-E\right]=0$ is an algebraic equation for $\beta$, and let $2M$ be the degree of the equation. The main result is that when the eigenvalue equation has solutions $\beta_i~(i=1,\cdots,2M)$ with $\left|\beta_1\right|\leq\left|\beta_2\right|\leq\cdots\leq\left|\beta_{2M-1}\right|\leq\left|\beta_{2M}\right|$, $C_\beta$ is given by the trajectory of $\beta_M$ and $\beta_{M+1}$ under a condition $\left|\beta_M\right|=\left|\beta_{M+1}\right|$. It is obtained as the condition to construct continuum bands, which reproduce band structure for a large crystal with open boundaries. We note that in Hermitian systems, this condition reduces to $C_\beta:\left|\beta\right|=1$, meaning that $k$ becomes real. In previous works, systems with $M=1$ have been studied in general cases~\cite{Yao2018} and in limited cases~\cite{Kunst2019}.

A byproduct of our theory is that one can prove the bulk-edge correspondence. The bulk-edge correspondence has been discussed, but in most cases, it has not been shown rigorously but by observation on some particular cases, together with an analogy to Hermitian systems. It in fact shows that the bulk-edge correspondence for the real Bloch wave vector cannot be true in non-Hermitian systems. In this paper, we show the bulk-edge correspondence in the non-Hermitian Su-Schrieffer-Heeger (SSH) model with the generalized Brillouin zone and discuss the relationship between a topological invariant in the bulk and existence of the edge states.

%
%

We start with a 1D tight-binding model, with its Hamiltonian given by
\begin{equation}
H=\sum_n\sum_{i=-N}^N\sum_{\mu,\nu=1}^qt_{i,\mu\nu}c_{n+i,\mu}^\dag c_{n,\nu},
\label{eq1}
\end{equation}
where $N$ represents the range of the hopping and $q$ represents the degrees of freedom per unit cell. This Hamiltonian can be non-Hermitian, meaning that $t_{i,\mu\nu}$ is not necessarily equal to $t_{-i,\nu\mu}^\ast$. Then one can write the real-space eigen-equation as $H\ket{\psi}=E\ket{\psi}$, where the eigenvector is written as $\ket{\psi}=\left(\cdots,\psi_{1,1},\cdots,\psi_{1,q},\psi_{2,1},\cdots,\psi_{2,q},\cdots\right)^{\rm T}$ in an open chain. Thanks to the spatial periodicity, one can write the eigenvector as a linear combination:
\begin{equation}
\psi_{n,\mu}=\sum_j\phi_{n,\mu}^{\left(j\right)},~\phi_{n,\mu}^{\left(j\right)}=\left(\beta_j\right)^n\phi_\mu^{\left(j\right)},~(\mu=1,\cdots,q).
\label{eq2}
\end{equation}
By imposing that $\phi^{\left(j\right)}_{n,\mu}$ is an eigenstate, one can obtain the eigenvalue equation (for example, see Eq.~(\ref{eq7})) for $\beta=\beta_j$ as
\begin{equation}
\det\left[{\cal H}\left(\beta\right)-E\right]=0.
\label{eq3}
\end{equation}
Here this eigenvalue equation is an algebraic equation for $\beta$ with an even degree $2M$ in general cases~\cite{SM}.

%
%

\begin{figure}[]
\includegraphics[width=8.5cm]{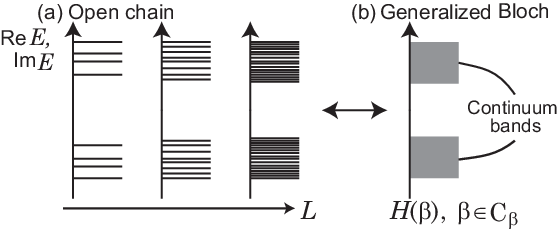}
\caption{\label{fig:band}Schematic figure of the band structure (a) in a finite open chain with various system sizes $L$, and (b) in the generalized Bloch Hamiltonian. The vertical axis represents the distribution of the complex energy $E$.}
\end{figure}
One can see from Eq.~(\ref{eq2}) that $\beta$ corresponds to the Bloch wave number $k\in{\mathbb R}$ via $\beta={\rm e}^{ik}$ in Hermitian systems. The bulk-band structure for reality of $k$ reproduces the band structure of a long open chain. When extending this idea to non-Hermitian systems, we should choose values of $\beta$ such that the bands of the Hamiltonian ${\cal H}\left(\beta\right)$ reproduce those of a long open chain (Fig.~\ref{fig:band}). The levels are discrete in a finite open chain, and as the system size becomes larger, the levels become dense and the asymptotically form continuum bands (Fig.~\ref{fig:band}). Therefore, in order to find the generalized Brillouin zone $C_\beta$, one should consider asymptotic behavior of level distributions in an open chain in the limit of a large system size. In Hermitian systems, $\left|\beta\right|$  is equal to unity, meaning that the eigenstates extend over the bulk. On the other hand, in non-Hermitian systems, $\left|\beta\right|$ is not necessarily unity, and these states may be localized at either end of the chain. Therefore, these bands cannot be called bulk bands, but should be called continuum bands. These states are incompatible with the periodic boundaries. The continuum bands are formed by changing $\beta$ continuously along $C_\beta$, as we show later. 

%
%

Next. we find how to determine the generalized Brillouin zone $C_\beta$, which determines the continuum bands. Here we number the solutions $\beta_i~(i=1,\cdots,2M)$ of Eq.~(\ref{eq3}) so as to satisfy $\left|\beta_1\right|\leq\left|\beta_2\right|\leq\cdots\leq\left|\beta_{2M-1}\right|\leq\left|\beta_{2M}\right|$. We find that the condition to get the continuum bands can be written as
\begin{equation}
\left|\beta_M\right|=\left|\beta_{M+1}\right|, 
\label{eq4}
\end{equation}
and the trajectory of $\beta_M$ and $\beta_{M+1}$ gives $C_\beta$. In Hermitian systems, we can prove that Eq.~(\ref{eq4}) becomes $\left|\beta_M\right|=\left|\beta_{M+1}\right|=1$~\cite{SM}, and $C_\beta$ is a unit circle, $\left|\beta\right|=1$. When $M=1$,
this condition physically corresponds to a condition for the formation of a standing wave in an open chain as proposed in Ref.~\onlinecite{Yao2018}. We discuss this point in Sec.~S\Rnum{1} in the Supplemental Material~\cite{SM}.
%

%
%

To get Eq.~(\ref{eq4}), we focus on boundary conditions in an open chain. Here we provide an outline of the process by which we arrive at Eq.~(\ref{eq4}), and we give a detailed discussion in Secs.~S\Rnum{2} and S\Rnum{3} in the Supplemental Material~\cite{SM}. We impose the wave function in Eq.~(\ref{eq2}) to represent an eigenstate. Apart from the positions near the two ends, it leads to the eigenvalue equation (\ref{eq3}). The boundary conditions place another constraint on the values of $\beta_i~(i=1,\cdots,2M)$ in the form of an algebraic equation. We now suppose the system size $L$ to be quite large and consider a condition to achieve densely distributed levels (Fig.~\ref{fig:band}). The equation consists of terms of the form $\left(\beta_{i_1}\beta_{i_2}\cdots\beta_{i_M}\right)^L$. When $\left|\beta_M\right|\neq\left|\beta_{M+1}\right|$, there is only one leading term proportional to $\left(\beta_{M+1}\cdots\beta_{2M}\right)^L$, which does not allow continuum bands. 
Only when $\left|\beta_M\right|=\left|\beta_{M+1}\right|$, are there two leading terms proportional to $\left(\beta_{M}\beta_{M+2}\cdots\beta_{2M}\right)^L$ and to $\left(\beta_{M+1}\beta_{M+2}\cdots\beta_{2M}\right)^L$. In such a case, the relative phase between $\beta_M$ and $\beta_{M+1}$ can be changed almost continuously for a large $L$, producing the continuum bands. 
We note that our condition Eq.~(\ref{eq4}) is independent of any boundary conditions. In Ref.~\onlinecite{Yao2018}, it was proposed that the continuum bands require $\left|\beta_i\right|=\left|\beta_j\right|$. Nonetheless, this is not sufficient; except for the case $\left|\beta_M\right|=\left|\beta_{M+1}\right|$, it does not allow the continuum bands.

%
%

\begin{figure}[]
\includegraphics[width=8.5cm]{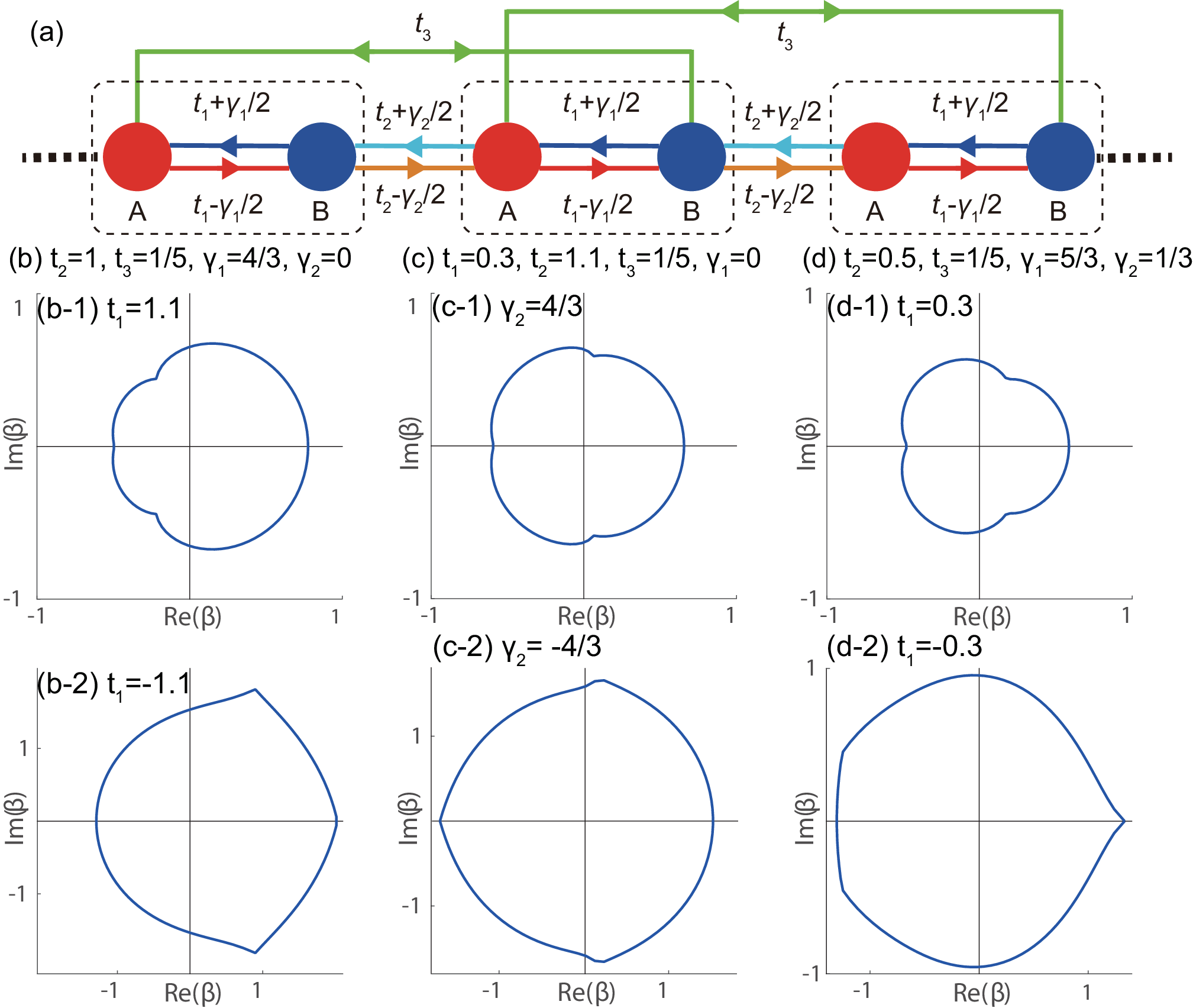}
\caption{\label{fig:SSH}(a) Non-Hermitian SSH model. The dotted boxes indicate the unit cell. (b)-(d) Generalized Brillouin zone $C_\beta$ of this model. The values of the parameters are (b) $t_2=1,t_3=1/5,\gamma_1=4/3$, and $\gamma_2=0$, with (b-1) $t_1=1.1$ and (b-2) $t_1=-1.1$; (c) $t_1=0.3,t_2=1.1,t_3=1/5$, and $\gamma_1=0$, with (c-1) $\gamma_2=4/3$ and (c-2) $\gamma_2=-4/3$; and (d) $t_2=0.5,t_3=1/5,\gamma_1=5/3$, and $\gamma_2=1/3$, with (d-1) $t_1=0.3$ and (d-2) $t_1=-0.3$.}
\end{figure}
We apply Eq.~(\ref{eq4}) to the non-Hermitian SSH model as shown in Fig.~\ref{fig:SSH}(a). 
It is given by
\begin{eqnarray}
H&=&\sum_n\left[\left(t_1+\frac{\gamma_1}{2}\right)c_{n,{\rm A}}^\dag c_{n,{\rm B}}+\left(t_1-\frac{\gamma_1}{2}\right)c_{n,{\rm B}}^\dag c_{n,{\rm A}}\right. \nonumber\\
&&+\left(t_2+\frac{\gamma_2}{2}\right)c_{n,{\rm B}}^\dag c_{n+1,{\rm A}}+\left(t_2-\frac{\gamma_2}{2}\right)c_{n+1,{\rm A}}^\dag c_{n,{\rm B}} \nonumber\\
&&\left.+t_3\left(c_{n,{\rm A}}^\dag c_{n+1,{\rm B}}+c_{n+1,{\rm B}}^\dag c_{n,{\rm A}}\right)\right],
\label{eq5}
\end{eqnarray}
where $t_1,t_2,t_3,\gamma_1$, and $\gamma_2$ are real. 
The generalized Bloch Hamiltonian ${\cal H}\left(\beta\right)$ can be obtained by a replacement ${\rm e}^{ik}\rightarrow\beta$, similarly to Hermitian systems, as ${\cal H}\left(\beta\right)=R_+\left(\beta\right)\sigma_++R_-\left(\beta\right)\sigma_-$, where $\sigma_\pm=\left(\sigma_x\pm i\sigma_y\right)/2$, and $R_\pm\left(\beta\right)$ are given by
\begin{eqnarray}
R_+\left(\beta\right)&=&\left(t_2-\frac{\gamma_2}{2}\right)\beta^{-1}+\left(t_1+\frac{\gamma_1}{2}\right)+t_3\beta, \nonumber\\
R_-\left(\beta\right)&=&t_3\beta^{-1}+\left(t_1-\frac{\gamma_1}{2}\right)+\left(t_2+\frac{\gamma_2}{2}\right)\beta.
\label{eq6}
\end{eqnarray}
Therefore the eigenvalue equation can be written as
\begin{equation}
R_+\left(\beta\right)R_-\left(\beta\right)=E^2,
\label{eq7}
\end{equation}
\noindent which is a quartic equation for $\beta$; i.e., $M=2$, having four solutions $\beta_i~(i=1,\cdots,4)$ satisfying $\left|\beta_1\right|\leq\left|\beta_2\right|\leq\left|\beta_3\right|\leq\left|\beta_4\right|$. Then Eq.~(\ref{eq4}) is given by $\left|\beta_2\right|=\left|\beta_3\right|$~\cite{SM}.

%
%

The trajectory of $\beta_2$ and $\beta_3$ satisfying the condition $\left|\beta_2\right|=\left|\beta_3\right|$ determines the generalized Brillouin zone $C_\beta$, and it is shown in Figs.~\ref{fig:SSH}(b)-\ref{fig:SSH}(d) for various values of the parameters. It always forms a loop enclosing the origin on the complex plane. Nonetheless, we do not have a rigorous proof that $C_{\beta}$ is always a single loop encircling the origin. 
We find some features of $C_\beta$. First, our result does not depend on whether $\left|\beta\right|$ is larger or smaller than unity, as opposed to the suggestions in previous works~\cite{Yao2018,Liu2019}; in Fig.~\ref{fig:SSH}(d-2), $\left|\beta\right|$ takes both values more than 1 and values less than 1. Second, $C_\beta$ can be a unit circle even for non-Hermitian cases; for example, when $t_1=t_3=\gamma_2=0$. Finally, $C_\beta$ can have cusps, corresponding to the cases where three solutions share the same absolute value~\cite{SM}.

%
%

\begin{figure}[]
\includegraphics[width=8.5cm]{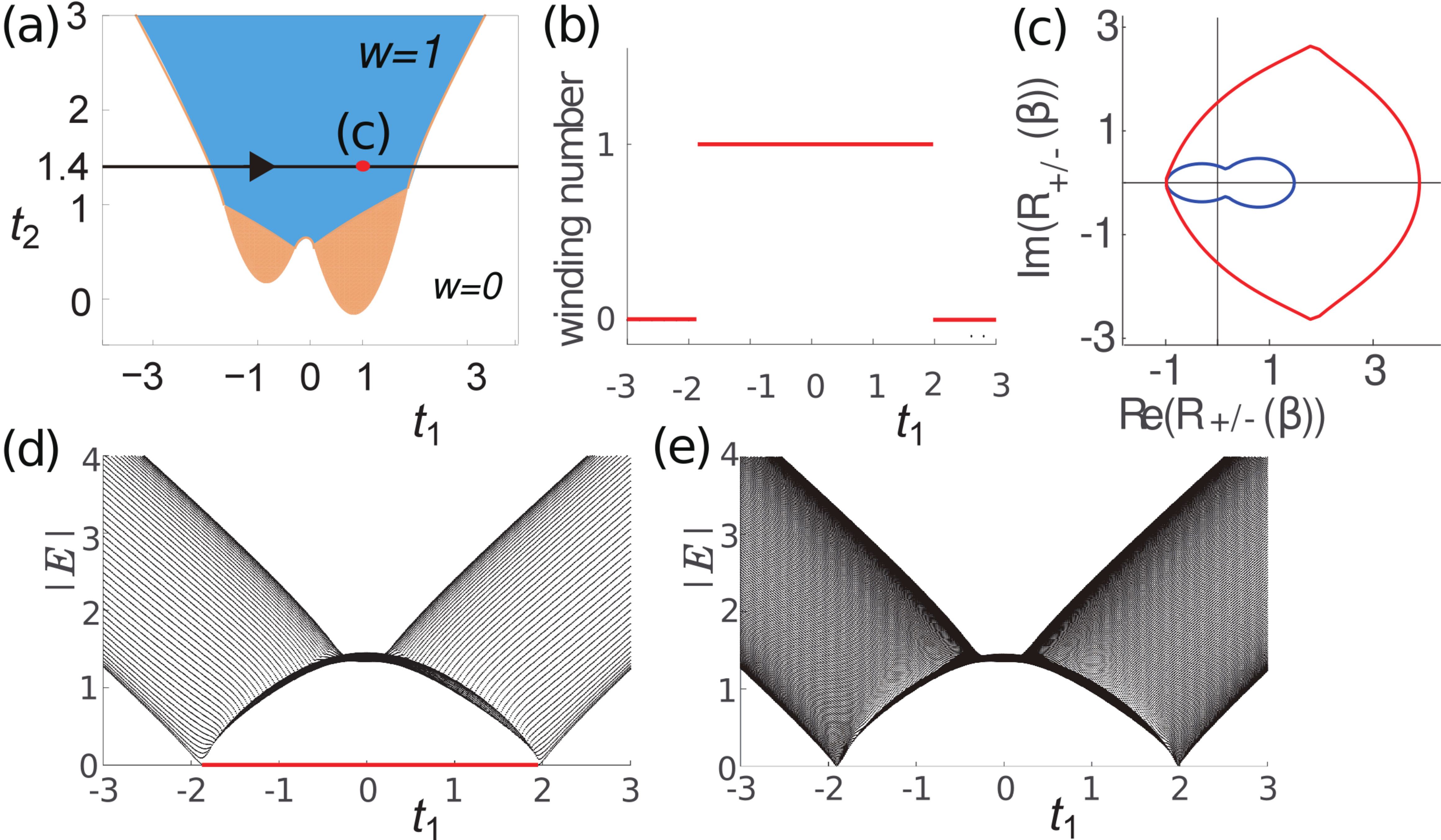}
\caption{\label{fig3}(Color online) Phase diagram and bulk-edge correspondence with $t_3=1/5,\gamma_1=5/3$, and $\gamma_2=1/3$. (a) Phase diagram on the $t_1$-$t_2$ plane. The blue region represents that the winding number is $1$, and the orange region represents that the system has exceptional points. Along the black arrow in (a) with $t_2=1.4$, we show the results for (b) the winding number, (d) energy bands in a finite open chain, and (e) the continuum bands from the generalized Brillouin zone $C_\beta$. The edge states are shown in red in (d). (c) shows $\ell_+$ (red) and $\ell_-$ (blue) on the $\bm{R}$ plane with $t_1=1$ and $t_2=1.4$.} 
\end{figure}
We calculate the winding number $w$ for the Hamiltonian ${\cal H}(\beta)$. Thanks to the chiral symmetry, $w$ can be defined as~\cite{SM}
\begin{equation}
w=-\frac{w_+-w_-}{2},~w_{\pm}=\frac{1}{2\pi}\left[\arg R_\pm\left(\beta\right)\right]_{C_\beta},
\label{eq8}
\end{equation}
where $\left[\arg R_\pm\left(\beta\right)\right]_{C_\beta}$ means the change of the phase of $R_\pm\left(\beta\right)$ as $\beta$ goes along the generalized Brillouin zone $C_\beta$ in a counterclockwise way. It is proposed that $w$ corresponds to the presence or absence of the topological edge states~\cite{Yao2018}. 

%
%

We show how the gap closes in our model. It closes when $E=0$, i.e., $R_+\left(\beta\right)=0$ or $R_-\left(\beta\right)=0$. Let $\beta=\beta_i^a~(i=1,2,~a=+,-)$ denote the solutions of the equation $R_{a}\left(\beta\right)=0$, with $\left|\beta_1^a\right|\leq\left|\beta_2^a\right|$. When $E=0$ is in the continuum bands, Eq.~(\ref{eq4}) should be satisfied for the four solutions $\beta_i^\pm~(i=1,2)$. It can be classified into two cases, (a) $\left|\beta_1^a\right|\leq\left|\beta_2^a\right|=\left|\beta_1^{-a}\right|\leq\left|\beta_2^{-a}\right|~(a=+,-)$, and (b) $\left|\beta_1^a\right|\leq\left|\beta_1^{-a}\right|=\left|\beta_2^{-a}\right|\leq\left|\beta_2^a\right|~(a=+,-)$. In case (a), as we change one parameter, the gap closes at $E=0$, and $w_+$ and $-w_-$ change by 1 at the same time, giving rise to the change of the winding number by unity. On the other hand, in case (b), only one of the two coefficients $R_\pm\left(\beta\right)$ becomes zero, and it represents an exceptional point.

%
%

\begin{figure}[]
\includegraphics[width=8.5cm]{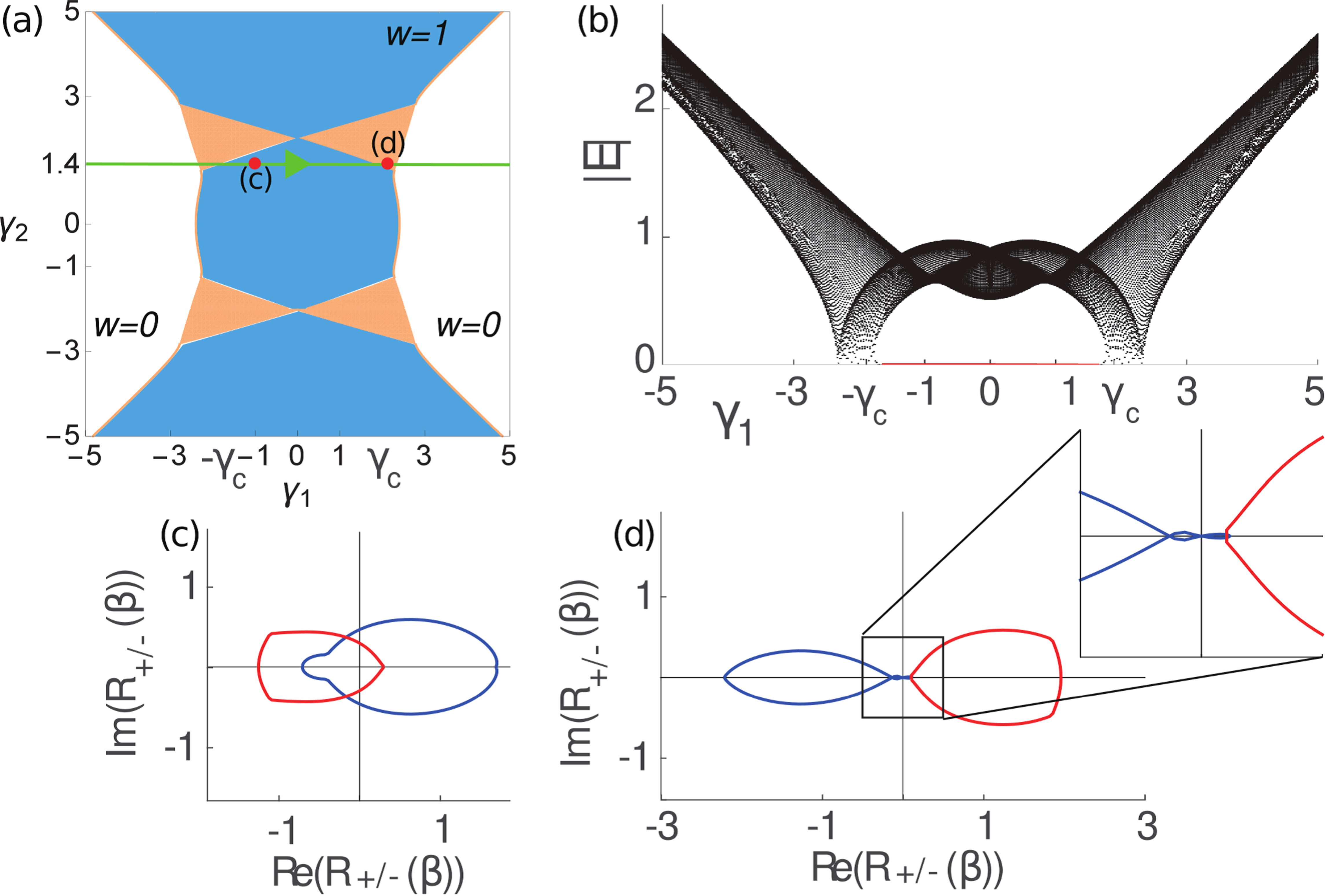}
\caption{\label{fig4}Phase diagram and bulk-edge correspondence with $t_1=0,t_2=1$, and $t_3=1/5$. (a) Phase diagram on the $\gamma_1$-$\gamma_2$ plane. The blue region represents that the winding number is $1$, and the orange region represents that the system has exceptional points. 
(b) Energy bands calculated along the green arrow in (a) with $\gamma_2=1.4$. Note that $\gamma_c\simeq1.89$. The edge states are shown in red. (c),(d) Loops $\ell_+$ (red) and $\ell_-$ (blue) on the ${\bm R}$ plane. The values of the parameters are (c) $\gamma_1=-1$ and $\gamma_2=1.4$, and (d) $\gamma_1=2.1$ and $\gamma_2=1.4$. Note that $\ell_-$ passes the origin in (d), which corresponds to exceptional points.}
\end{figure}

We obtain the phase diagram on the $t_1$-$t_2$ plane in Fig.~\ref{fig3}(a) and on the $\gamma_1$-$\gamma_2$ plane in Fig.~\ref{fig4}(a). In these phase diagrams, the winding number $w$ is $1$ in the blue region. By definition, $w$ changes only when $R_\pm\left(\beta\right)=0$ on the generalized Brillouin zone $C_\beta$, and the gap closes. The energy bands in a finite open chain calculated along the black arrow in Fig.~\ref{fig3}(a) are shown in Fig.~\ref{fig3}(d), and one can confirm that the edge states appear in the region where $w=1$. In addition, the continuum bands using $C_\beta$ (Fig.~\ref{fig3}~(e)) agree with these energy bands. 
In Fig.~\ref{fig4}(b), we give the energy bands calculated along the green arrow in Fig.~\ref{fig4}(a), and the edge states appear similarly to Fig.~\ref{fig3}(d). On the other hand, the system has the exceptional points in the orange region. The phase with the exceptional points extends over a finite region~\cite{SM}.

%
%

We discuss the bulk-edge correspondence in our model. The loops $\ell_{\pm}$ drawn by $R_{\pm}\left(\beta\right)$ on the ${\bm R}$ plane are shown in Fig.~\ref{fig3}(c) and in Figs.~\ref{fig4}(c) and \ref{fig4}(d) for certain values of the parameters. In both Fig.~\ref{fig3}(c) and Fig.~\ref{fig4}(c), the system has the winding number $w=1$, since both $\ell_+$ and $\ell_-$ surround the origin ${\cal O}$, leading to $w_+=-1$ and $w_-=1$. 
In Fig.~\ref{fig4}(a), one can continuously change the values of the parameters to the Hermitian limit, $\gamma_1,\gamma_2\rightarrow0$, while keeping the gap open and while $w=1$ remains. The same is true for Fig.~\ref{fig3}(a). Therefore, by following the proof in Hermitian cases~\cite{Ryu2002}, one can prove the bulk-edge correspondence even for the non-Hermitian cases, and the existence of zero-energy states is derived~\cite{SM}. On the other hand, $\ell_-$ passes ${\cal O}$ as shown in Fig.~\ref{fig4}(d), where the system has exceptional points. We note that the winding number is not well defined in this case.

%
%

In summary, we establish a generalized Bloch band theory in 1D tight-binding systems and obtain the condition for the continuum bands. We show the way to construct the generalized Brillouin zone $C_\beta$, which is fundamental for obtaining the continuum bands. Here the Bloch wave number $k$ takes complex values in non-Hermitian systems. 
Our conclusion, $\left|\beta_M\right|=\left|\beta_{M+1}\right|$, is physically reasonable in several aspects. First, it is independent of any boundary conditions. Thus, for a long open chain, irrespective of any boundary conditions, the spectrum asymptotically approaches the same continuum bands calculated from $C_\beta$~\cite{SM}. Second, it reproduces the known result in the Hermitian limit, i.e., $\left|\beta\right|=1$. Third, the form of the condition is invariant under the replacement $\beta\rightarrow1/\beta$. Suppose the numbering of the sites is reversed by setting $n^\prime=L+1-n$ for the site index $n(=1,\cdots,L)$; then $\beta$ becomes $\beta^\prime=1/\beta$, but the form of the condition is invariant: $\left|\beta_M^\prime\right|=\left|\beta_{M+1}^\prime\right|$.

%
%

Through this definition of the continuum bands, one can show the bulk-edge correspondence without ambiguity by defining the winding number $w$ from the generalized Brillouin zone in 1D systems with chiral symmetry. Indeed, we showed that the zero-energy states appear in the non-Hermitian SSH model when $w$ takes nonzero values, and we also revealed that these states correspond to topological edge states. It is left for future works to determine how to calculate the continuum bands for systems with other symmetries.

%
%

The construction of the generalized Brillouin zone can be extended to higher dimensions as well. In two-dimensional (2D) systems, we introduce the two parameters $\beta^x\left(={\rm e}^{ik_x}\right)$ and $\beta^y\left(={\rm e}^{ik_y}\right)$. Then the eigenvalue equation $\det\left[{\cal H}\left(\beta^x,\beta^y\right)-E\right]=0$, where ${\cal H}\left(\beta^x,\beta^y\right)$ is a 2D generalized Bloch Hamiltonian, is an algebraic equation for $\beta^x$ and $\beta^y$. If we fix $\beta^y$ $\left(\beta^x\right)$, this system can be regarded as a 1D system, and the criterion is given by $\left|\beta^x_{M_x}\right|=\left|\beta^x_{M_x+1}\right|$ $\left(\left|\beta^y_{M_y}\right|=\left|\beta^y_{M_y+1}\right|\right)$, where $2M_x$ $\left(2M_y\right)$ is the degree of the eigenvalue equation for $\beta^x$ $\left(\beta^y\right)$. 
Thus, we can get the conditions for the continuum bands. Nevertheless, it is still an open question how to determine the generalized Brillouin zone in higher dimensions.

%
%

We also apply our theory to the tight-binding model proposed in Ref.~\onlinecite{Lee2016}, and we show that the Bloch wave number $k$ has a nonzero imaginary part, and the bulk-edge correspondence can be established with $k\in{\mathbb C}$~\cite{SM}. We conclude that some previous works on the bulk-edge correspondence using the reality of the Bloch wave vector require further investigation.

%
%

This work was supported by a Grant-in-Aid for Scientific Research (Grants No.~JP18H03678 and No.~JP16J07354) by MEXT, Japan; by CREST, JST (No.~JP-MJCR14F1); and by the MEXT Elements Strategy Initiative to Form Core Research Center (TIES). K. Y. was also supported by JSPS KAKENHI (Grant No.~18J22113).
%

\providecommand{\noopsort}[1]{}
\end{document}


\preprint{APS/123-QED}

\title{Supplemental Material for ``Non-Bloch Band Theory of Non-Hermitian Systems"}

\author{Kazuki Yokomizo}
\affiliation{Department of Physics, Tokyo Institute of Technology, 2-12-1 Ookayama, Meguro-ku, Tokyo, 152-8551, Japan}
\author{Shuichi Murakami}
\affiliation{Department of Physics, Tokyo Institute of Technology, 2-12-1 Ookayama, Meguro-ku, Tokyo, 152-8551, Japan}
\affiliation{TIES, Tokyo Institute of Technology, 2-12-1 Ookayama, Meguro-ku, Tokyo, 152-8551, Japan}%




%
\pacs{Valid PACS appear here}
\maketitle
%
%

\section{\label{sec1}Simple non-Hermitian tight-binding model}
In this section, we study a simple non-Hermitian tight-binding model in terms of the non-Bloch band theory. The Hamiltonian of this system is given by
\begin{equation}
H=\sum_{n=1}^{L-1}\left(t_{\rm R}c_{n+1}^\dag c_n+t_{\rm L}c_{n}^\dag c_{n+1}\right),
\label{eq10}
\end{equation}
where $t_{\rm R},t_{\rm L}\in{\mathbb R}$ are the asymmetric hopping amplitudes, and $L$ is the system size. We put $t_{\rm R}$ and $t_{\rm L}$ to be positive, for simplicity. For the eigenvector $\ket{\psi}=\left(\psi_1,\cdots,\psi_L\right)^{\rm T}$, the real-space eigen-equation $H\ket{\psi}=E\ket{\psi}$ can be written as
\begin{equation}
t_{\rm R}\psi_{n-1}+t_{\rm L}\psi_{n+1}=E\psi_n,~(n=1,\cdots,L).
\label{eq20}
\end{equation}
Henceforth we imply $\psi_0=\psi_{L+1}=0$ as a boundary condition. From the theory of linear difference equations, a general solution of the recursion equation Eq.~(\ref{eq20}) is written as
\begin{equation}
\psi_n=\left(\beta_1\right)^n\phi^{\left(1\right)}+\left(\beta_2\right)^n\phi^{\left(2\right)},
\label{eq30}
\end{equation}
where $\beta_j~(j=1,2)$ are the solutions of the eigenvalue equation
\begin{equation}
t_{\rm R}\beta_j^{-1}+t_{\rm L}\beta_j=E.
\label{eq40}
\end{equation}
Here the boundary condition $\psi_0=0$ gives $\phi^{\left(1\right)}+\phi^{\left(2\right)}=0$. By combining this equation, the boundary condition $\psi_{L+1}=0$ gives $\left(\beta_1/\beta_2\right)^{L+1}=1$, and we can rewrite this equation as
\begin{equation}
\frac{\beta_1}{\beta_2}={\rm e}^{2i\theta_m},~\left(\theta_m=\frac{m\pi}{L+1},m=1,\cdots,L\right).
\label{eq50}
\end{equation}
Then we get
\begin{equation}
\left|\beta_1\right|=\left|\beta_2\right|,
\label{eq60}
\end{equation}
and we obtain $r\equiv\left|\beta_{1,2}\right|=\sqrt{t_{\rm R}/t_{\rm L}}$ since $\beta_1\beta_2=t_{\rm R}/t_{\rm L}$ from Eq.~(\ref{eq40}). Therefore $\beta_j~(j=1,2)$ can be given by
\begin{equation}
\beta_1^{\left(m\right)}=r{\rm e}^{i\theta_m},~\beta_2^{\left(m\right)}=r{\rm e}^{-i\theta_m},
\label{eq70}
\end{equation}
and the eigenstate Eq.~(\ref{eq30}) and the eigenenergy Eq.~(\ref{eq40}) can be written as
\begin{eqnarray}
\psi_n^{\left(m\right)}&\propto&\left(r{\rm e}^{i\theta_m}\right)^n-\left(r{\rm e}^{-i\theta_m}\right)^n\propto r^n\sin n\theta_m, \nonumber\\
E^{\left(m\right)}&=&2\sqrt{t_{\rm R}t_{\rm L}}\cos\theta_m.
\label{eq80}
\end{eqnarray}
Then we can get the discrete energy levels in a finite open chain.

As the system size $L$ becomes larger, these energy levels becomes dense, and finally, the energy spectrum can be obtained in the thermodynamic limit $L\rightarrow\infty$. These continuum bands at $L\rightarrow\infty$ can be obtained by setting Eq.~(\ref{eq60}). Namely Eq.~(\ref{eq60}) leads to
\begin{equation}
\beta_1=\sqrt{\frac{t_{\rm R}}{t_{\rm L}}}{\rm e}^{i\theta},~\beta_2=\sqrt{\frac{t_{\rm R}}{t_{\rm L}}}{\rm e}^{-i\theta},
\label{eq90}
\end{equation}
and from Eq.~(\ref{eq40}), we get
\begin{equation}
E=2\sqrt{t_{\rm R}t_{\rm L}}\cos\theta,
\label{eq100}
\end{equation}
\noindent giving the continuum bands by changing $\theta\in{\mathbb R}$. Therefore we can interpret Eq.~(\ref{eq60}) as the condition for the continuum bands. 

In Eq.~(\ref{eq80}), apart from the factor $r^n$, the wave function $\psi_n^{\left(m\right)}$ represents a standing wave, as a superposition of two counterpropagating plane waves. Thus the condition for the continuum bands Eq.~(\ref{eq60}) can be also understood as a condition for formation of a standing wave in this system.

%
%

\section{\label{sec2}Non-Bloch band theory}
In this section, we describe a systematic way to get the generalized Brillouin zone $C_\beta$ in general cases. In particular, $C_\beta$ becomes a unit circle in Hermitian cases.
\subsection{\label{sec2-1}General recipe for deriving the generalized Brillouin zone}
\begin{figure}[]
\includegraphics[width=8.5cm]{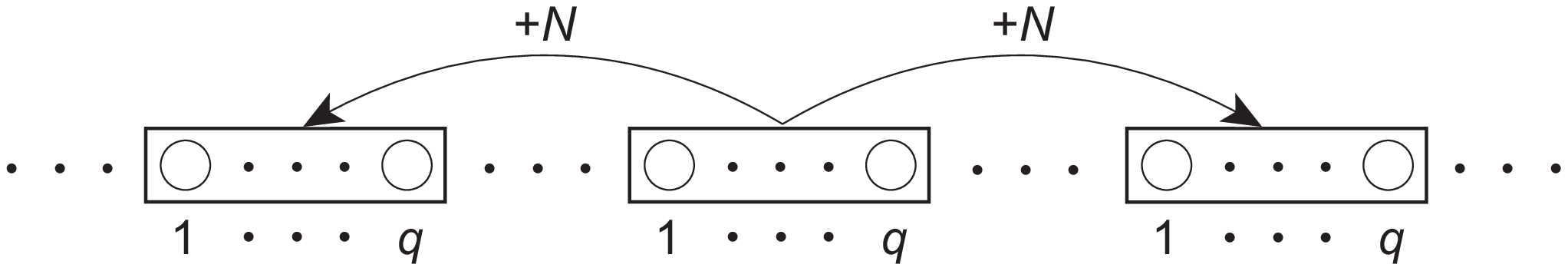}
\caption{\label{fig0}Schematic figure of a 1D tight-binding system. A unit cell includes $q$ degrees of freedom, and the range of hopping is $N$.}
\end{figure}
We start with a one-dimensional (1D) tight-binding model in an open chain as shown in Fig.~\ref{fig0}. A unit cell is composed of $q$ degrees of freedom, such as sublattices, spins, or orbitals, and the range of hopping is $N$, meaning that electrons hop up to the $N$-th nearest unit cells. Then its Hamiltonian can be written as
\begin{equation}
H=\sum_n\sum_{i=-N}^N\sum_{\mu,\nu=1}^qt_{i,\mu\nu}c_{n+i,\mu}^\dag c_{n,\nu},
\label{eq11a}
\end{equation}
where $c_{n,\mu}^\dag~\left(c_{n,\mu}\right)$ is a creation (an annihilation) operator of an electron with index $\mu~(\mu=1,\cdots,q)$ in the $n$-th unit cell, and $t_{i,\mu\nu}$ is a hopping amplitude to the $i$-th nearest unit cell. This Hamiltonian can be non-Hermitian, meaning that $t_{i,\mu\nu}$ is not necessarily equal to $t_{-i,\nu\mu}^\ast$. Here one can write the real-space eigen-equation as $H\ket{\psi}=E\ket{\psi}$, where the eigenvector is written as $\ket{\psi}=\left(\cdots,\psi_{1,1},\cdots,\psi_{1,q},\psi_{2,1},\cdots,\psi_{2,q},\cdots\right)^{\rm T}$. Thanks to the spatial periodicity, we can write the eigenvector as a linear combination:
\begin{equation}
\psi_{n,\mu}=\sum_j\phi_{n,\mu}^{\left(j\right)},~\phi_{n,\mu}^{\left(j\right)}=\left(\beta_j\right)^n\phi_\mu^{\left(j\right)},~(\mu=1,\cdots,q).
\label{eq21a}
\end{equation}
Then, by imposing that $\phi^{\left(j\right)}_{n,\mu}$ is an eigenstate, the eigen-equation can be rewritten for $\beta=\beta_j$ as
\begin{eqnarray}
\sum_{\nu=1}^q\left[{\cal H}\left(\beta\right)\right]_{\mu\nu}\phi_\nu=E\phi_\mu,~\left(\mu=1,\cdots,q\right),
\label{eq31a}
\end{eqnarray}
where a generalized Bloch Hamiltonian ${\cal H}\left(\beta\right)$ is given by
\begin{equation}
\left[{\cal H}\left(\beta\right)\right]_{\mu\nu}=\sum_{i=-N}^Nt_{i,\mu\nu}\beta^i,~\left(\mu,\nu=1,\cdots,q\right).
\label{eq41a}
\end{equation}
We note that in Hermitian cases, Eq.~(\ref{eq31a}) is nothing but the eigen-equation for the Bloch eigenstates. Therefore we can get the eigenvalue equation as
\begin{equation}
\det\left[{\cal H}\left(\beta\right)-E\right]=0.
\label{eq51a}
\end{equation}
We note that the eigenvalue equation is an algebraic equation for $\beta$ with an even degree $2M=2qN$ in general cases, where $M=qN$ because each element of the $q\times q$ matrix ${\cal H}\left(\beta\right)$ has the form of Eq.~(\ref{eq41a}).

Next we show how to determine the generalized Brillouin zone $C_\beta$, which determines continuum bands. It is one of the main results of our work. Let $\beta_j~(j=1,\cdots,2M)$ be the solutions of the eigenvalue equation Eq.~(\ref{eq51a}). Here we number the $2M$ solutions so as to satisfy
\begin{equation}
\left|\beta_1\right|\leq\left|\beta_2\right|\leq\cdots\leq\left|\beta_{2M-1}\right|\leq\left|\beta_{2M}\right|. 
\label{eq61a}
\end{equation}
We find that the condition to get continuum bands can be written as
\begin{equation}
\left|\beta_M\right|=\left|\beta_{M+1}\right|,
\label{eq71a}
\end{equation}
and then the trajectory of $\beta_M$ and $\beta_{M+1}$ satisfying Eq.~(\ref{eq71a}) determines $C_\beta$. The derivation of Eq.~(\ref{eq71a}) is given in Sec.~\ref{sec3}. For example, in the non-Hermitian SSH model, $C_\beta$ is shown in Fig.~\ref{fig:BZ-SM} (a). We note that Eq.~(\ref{eq71a}) was proposed in Ref.~\onlinecite{Yao2018} in a simpler case of Eq.~(\ref{eq51a}) being a quadratic equation for $\beta$ corresponding to $M=1$, but for general cases its condition has not been known so far. We also mention Ref.~\onlinecite{Kunst2019}, which discusses the condition for continuum bands for limited cases of $M=1$, $J_L=J_R=J$, and $J^2=0$, where $J_L$ and $J_R$ are matrices formed by $t_{1,\mu\nu}$ and $t_{-1,\mu\nu}$ in our notation.

We here mention special cases in which the degree of the eigenvalue equation Eq.~(\ref{eq51a}) is less than $2M=2qN$. This occurs when some hopping amplitudes are zero. Even in these cases, one can restore the eigenvalue equation to an algebraic equation with the degree $2M$ for $\beta$, by including all hopping amplitudes to the $N$-th nearest unit cells to be non-zero, through an addition of infinitesimally small hopping amplitudes to the otherwise zero hopping amplitudes. Then Eq.~(\ref{eq71a}) can be applied in this situation. By taking the zero limit of all these infinitesimal hopping amplitudes, continuum bands in the original system can be obtained. In Sec.~\ref{sec4-2}, we describe such special cases, and we show how to apply our theory to them.
\subsection{\label{sec2-2}Hermitian cases}
In this subsection, we prove that the generalized Brillouin zone $C_\beta$ becomes a unit circle in Hermitian cases. In the following, we assume that the eigenvalue equation Eq.~(\ref{eq51a}) is an algebraic equation for $\beta$ of $2M$-th degree. In this case, it can be written as
\begin{equation}
\sum_{i=-M}^Ma_i\beta^i=0,
\label{eq11b}
\end{equation}
where $a_i~(i=-M,\cdots,M)$ are coefficients satisfying $a_{-i}=a_i^\ast$ and $a_0$ is real. These coefficients are functions of the eigenenergy $E$ and the hopping terms $t_{i,\mu\nu}$ included in Eq.~(\ref{eq11a}). Here the eigenenergy $E$ is real due to Hermiticity of the Hamiltonian. Equation~(\ref{eq11b}) has $2M$ solutions, and they are numbered so as to satisfy Eq.~(\ref{eq61a}). Since one can rewrite Eq.~(\ref{eq11b}) as
\begin{equation}
\sum_{i=-M}^M\left(a_{-i}\right)^\ast\left(\beta^\ast\right)^{-i}=
\sum_{i=-M}^M a_{i}\left(\beta^\ast\right)^{-i}=0,
\label{eq21b}
\end{equation}
the solutions always appear in pairs: $\left(\beta,1/\beta^\ast\right)$, and we get $\beta_{2M+1-j}=1/\beta_j^\ast$. In particular, one can get the relationship between $\beta_M$ and $\beta_{M+1}$ as $\beta_{M+1}=1/\beta_{M}^\ast$. Therefore Eq.~(\ref{eq71a}) can be rewritten as
\begin{equation}
\left|\beta_M\right|=\left|\beta_{M+1}\right|=1.
\label{eq31b}
\end{equation}
Thus we can conclude that $C_\beta$ becomes a unit circle in Hermitian systems.

%
%

\section{\label{sec3}Derivation of the condition for continuum bands}
In this section, we derive the condition for continuum bands Eq.~(\ref{eq71a}). To this end, we focus on boundary conditions in a finite open chain with $L$ unit cells. The wave functions are written in the form Eq.~(\ref{eq21a}), i.e.
\begin{equation}
\psi_{n,\mu}=\sum_{j=1}^{2M}\left(\beta_j\right)^n\phi_\mu^{\left(j\right)},~(n=1,\cdots L,~\mu=1,\cdots,q).
\label{eq12}
\end{equation}
The equations from a boundary condition include the $2qM$ unknown variables $\phi_\mu^{\left(j\right)}$. The real-space eigen-equation $H\ket{\psi}=E\ket{\psi}$ fixes the ratio between the values of $\phi_\mu^{\left(j\right)}$ sharing the same value of $j$. Therefore one can reduce the $2qM$ variables to the $2M$ variables $\phi_\mu^{\left(j\right)}$ with a single value of $\mu$, e.g. $\mu=1$. As a result, we can get $M$ equations for the $2M$ variables $\phi_1^{\left(j\right)}~(j=1,\cdots,2M)$ at the left end of the open chain around $n=1$:
\begin{equation}
\sum_{j=1}^{2M}f_i\left(\beta_j,E,{\cal S}\right)\phi^{\left(j\right)}_1=0,~\left(i=1,\cdots,M\right),
\label{eq22}
\end{equation}
and $M$ equations at the right end of the open chain around $n=L$:
\begin{equation}
\sum_{j=1}^{2M}g_i\left(\beta_j,E,{\cal S}\right)\left(\beta_j\right)^L\phi^{\left(j\right)}_1=0,~\left(i=1,\cdots,M\right),
\label{eq32}
\end{equation}
where ${\cal S}$ is the set of the system parameters $t_{i,\mu\nu}$. Here $f_i$ and $g_i$ are functions of $\beta_j,E$, and ${\cal S}$, and they do not depend on $L$. We emphasize here that any kinds of boundary conditions for an open chain can be written as $M$ equations (Eq.~(\ref{eq22})) for the left end and $M$ equations (Eq.~(\ref{eq32})) for the right end. By combining Eqs.~(\ref{eq22}) and (\ref{eq32}), we can obtain a condition for the existence of nontrivial solutions for $\phi_1^{\left(1\right)},\cdots,\phi_1^{\left(2M\right)}$:
\begin{eqnarray}
\left| \begin{array}{ccc}
f_1\left(\beta_1,E,{\cal S}\right)                       & \cdots & f_1\left(\beta_{2M},E,{\cal S}\right)                          \vspace{3pt}\\
\vdots                                                   & \vdots & \vdots                                                         \vspace{3pt}\\
f_M\left(\beta_1,E,{\cal S}\right)                       & \cdots & f_M\left(\beta_{2M},E,{\cal S}\right)                          \vspace{3pt}\\
g_1\left(\beta_1,E,{\cal S}\right)\left(\beta_1\right)^L & \cdots & g_1\left(\beta_{2M},E,{\cal S}\right)\left(\beta_{2M}\right)^L \vspace{3pt}\\
\vdots                                                       & \vdots & \vdots                                                     \vspace{3pt}\\
g_M\left(\beta_1,E,{\cal S}\right)\left(\beta_1\right)^L & \cdots & g_M\left(\beta_{2M},E,{\cal S}\right)\left(\beta_{2M}\right)^L
\end{array}\right|=0. \nonumber\\
\label{eq42}
\end{eqnarray}
We note that both $f_i$ and $g_i$ depend on boundary conditions but not on $L$.

The determinant included in Eq.~(\ref{eq42}) is an algebraic equation for $\beta_j~(j=1,\cdots,2M)$. By solving Eq.~(\ref{eq42}), one can calculate eigenenergies of the system with open boundaries. One cannot analytically solve these equations for a general system size. Nonetheless our aim is to see how the solutions for a large system size form continuum bands. Therefore we suppose $L$ to be quite large, and consider a condition to form densely distributed energy levels. Here Eq.~(\ref{eq42}) is expressed in a form
\begin{eqnarray}
\sum_{P,Q}F\left(\beta_{i\in P},\beta_{j\in Q},E,{\cal S}\right)\prod_{k\in P}\left(\beta_k\right)^L=0.
\label{eq52}
\end{eqnarray}
The sets $P$ and $Q$ are two disjoint subsets of the set $\{1,\cdots,2M\}$ such that the number of elements of each subset is $M$. In this way, the sum included in Eq.~(\ref{eq52}) is taken over all the sets $P$ and $Q$.

We now consider asymptotic behavior of the solutions of Eq.~(\ref{eq52}) for large $L$. When $\left|\beta_M\right|\neq\left|\beta_{M+1}\right|$, there is only one leading term proportional to $\left(\beta_{M+1}\cdots\beta_{2M}\right)^L$ in Eq.~(\ref{eq52}) in the limit of a large $L$. Thus it leads to the equation
\begin{equation}
F\left(\beta_{i\in P^\prime},\beta_{j\in Q^\prime},E,{\cal S}\right)=0
\label{eq62}
\end{equation}
with $P^\prime=\left\{M+1,\cdots,2M\right\}$ and $Q^\prime=\left\{1,\cdots,M\right\}$. This equation does not depend on $L$, and does not allow continuum bands. 
On the other hand, when $\left|\beta_M\right|=\left|\beta_{M+1}\right|$, there are two leading terms proportional to $\left(\beta_{M}\beta_{M+2}\cdots\beta_{2M}\right)^L$ and to $\left(\beta_{M+1}\beta_{M+2}\cdots\beta_{2M}\right)^L$, and the equation is written as
\begin{eqnarray}
\left(\frac{\beta_M}{\beta_{M+1}}\right)^L=-\frac{F\left(\beta_{i\in P_0},\beta_{j\in Q_0},E,{\cal S}\right)}{F\left(\beta_{i\in P_1},\beta_{j\in Q_1},E,{\cal S}\right)}
\label{eq72}
\end{eqnarray}
with $P_0=\left\{M+1,\cdots,2M\right\}$, $Q_0=\left\{1,\cdots,M\right\}$, $P_1=\left\{M,M+2,\cdots,2M\right\}$, and $Q_1=\left\{1,\cdots,M-1,M+1\right\}$. In such a case, we can expect that the relative phase between $\beta_M$ and $\beta_{M+1}$ can be changed almost continuously for a large $L$, producing the continuum bands. 

In Sec.~\ref{sec4}, we derive Eq.~(\ref{eq71a}) for the non-Hermitian SSH model as an example. In this model, Eq.~(\ref{eq52}) coming from the boundary condition is given by Eq.~(\ref{eq63b}).

%
%

\section{\label{sec4}Non-Hermitian SSH model}
In this section, we show how to get the eigenvalue equation and the condition for the continuum bands in the non-Hermitian SSH model. We show that even when an arbitrary complex potential is added on the boundary sites, the condition for the continuum bands Eq.~(\ref{eq71a}) is unchanged, and namely the generalized Brillouin zone is unchanged. Thus we emphasize that the condition for continuum bands, Eq.~(\ref{eq71a}), is independent of any boundary conditions.
\subsection{\label{sec4-1}Eigenvalue equation}
\begin{figure}[]
\includegraphics[width=8.5cm]{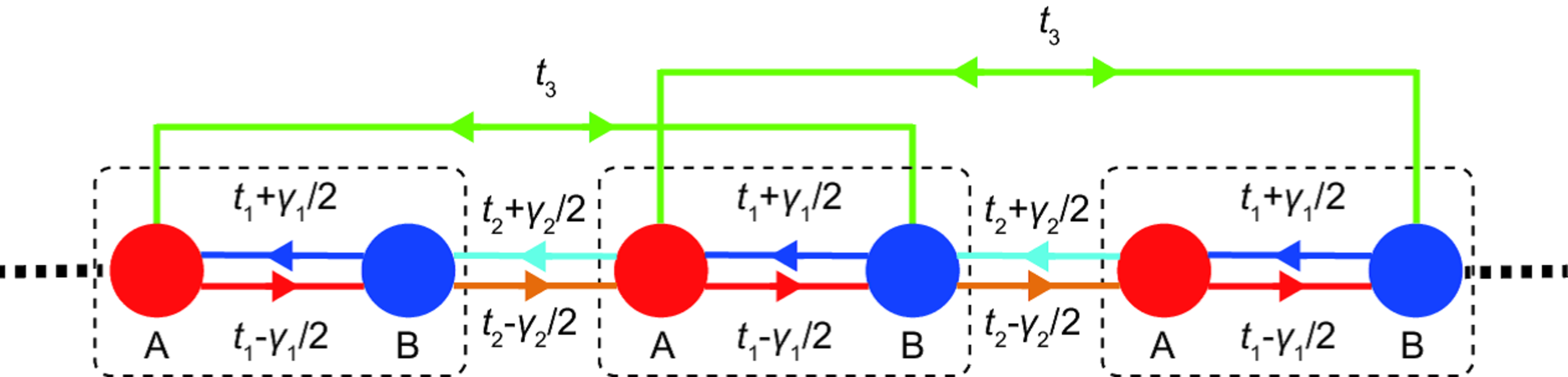}
\caption{\label{fig1a}Schematic figure of the non-Hermitian SSH model. The dotted boxes indicate the unit cell.}
\end{figure}
In this subsection, we derive the eigenvalue equation in the non-Hermitian SSH model as shown in Fig.~\ref{fig1a}. This model is the case of $q=2$ and $N=1$, giving $M=2$. The Hamiltonian can be written as
\begin{eqnarray}
H&=&\sum_n\left[\left(t_1+\frac{\gamma_1}{2}\right)c_{n,{\rm A}}^\dag c_{n,{\rm B}}+\left(t_1-\frac{\gamma_1}{2}\right)c_{n,{\rm B}}^\dag c_{n,{\rm A}}\right. \nonumber\\
&&+\left(t_2+\frac{\gamma_2}{2}\right)c_{n,{\rm B}}^\dag c_{n+1,{\rm A}}+\left(t_2-\frac{\gamma_2}{2}\right)c_{n+1,{\rm A}}^\dag c_{n,{\rm B}} \nonumber\\
&&\left.+t_3\left(c_{n,{\rm A}}^\dag c_{n+1,{\rm B}}+c_{n+1,{\rm B}}^\dag c_{n,{\rm A}}\right)\right],
\label{eq13a}
\end{eqnarray}
where $t_1,t_2,t_3,\gamma_1$, and $\gamma_2$ are real. Eigenvectors are written as $\ket{\psi}=\left(\cdots,\psi_{1,{\rm A}},\psi_{1,{\rm B}},\psi_{2,{\rm A}},\psi_{2,{\rm B}},\cdots\right)^{\rm T}$. Then we can explicitly write the equation for $\psi_{n,\mu}$ as
\begin{eqnarray}
\left(t_2-\frac{\gamma_2}{2}\right)\psi_{n-1,{\rm B}}+\left(t_1+\frac{\gamma_1}{2}\right)\psi_{n,{\rm B}}+t_3\psi_{n+1,{\rm B}}&=&E\psi_{n,{\rm A}}, \nonumber\\
t_3\psi_{n-1,{\rm A}}+\left(t_1-\frac{\gamma_1}{2}\right)\psi_{n,{\rm A}}+\left(t_2+\frac{\gamma_2}{2}\right)\psi_{n+1,{\rm A}}&=&E\psi_{n,{\rm B}}. \nonumber\\
\label{eq23a}
\end{eqnarray}
Here we take an ansatz for the wave function as a linear combination:
\begin{equation}
\psi_{n,\mu}=\sum_j\phi_{n,\mu}^{\left(j\right)},~(\mu={\rm A},{\rm B}),
\label{eq33a}
\end{equation}
where $\phi_{n,{\rm A}}^{\left(j\right)}$ and $\phi_{n,{\rm B}}^{\left(j\right)}$ takes the exponential form
\begin{equation}
\left(\phi_{n,{\rm A}}^{\left(j\right)},\phi_{n,{\rm B}}^{\left(j\right)}\right)=\left(\beta_j\right)^n\left(\phi_{\rm A}^{\left(j\right)},\phi_{\rm B}^{\left(j\right)}\right).
\label{eq43a}
\end{equation}
By substituting the exponential form Eq.~(\ref{eq43a}) into Eq.~(\ref{eq23a}), one can obtain the bulk eigen-equations:
\begin{eqnarray}
\left[\left(t_2-\frac{\gamma_2}{2}\right)\beta^{-1}+\left(t_1+\frac{\gamma_1}{2}\right)+t_3\beta\right]\phi_{\rm B}&=&E\phi_{\rm A}, \nonumber\\
\left[t_3\beta^{-1}+\left(t_1-\frac{\gamma_1}{2}\right)+\left(t_2+\frac{\gamma_2}{2}\right)\beta\right]\phi_{\rm A}&=&E\phi_{\rm B}, \nonumber\\
\label{eq53a}
\end{eqnarray}
and the generalized Bloch Hamiltonian ${\cal H}\left(\beta\right)$:
\begin{eqnarray}
{\cal H}\left(\beta\right)&=&R_+\left(\beta\right)\sigma_++R_-\left(\beta\right)\sigma_-, \nonumber\\
R_+\left(\beta\right)&=&\left(t_2-\gamma_2/2\right)\beta^{-1}+\left(t_1+\gamma_1/2\right)+t_3\beta,  \nonumber\\
R_-\left(\beta\right)&=&t_3\beta^{-1}+\left(t_1-\gamma_1/2\right)+\left(t_2+\gamma_2/2\right)\beta^{-1}, \nonumber\\
\label{eq63a}
\end{eqnarray}
where $\sigma_\pm=\left(\sigma_x\pm i\sigma_y\right)/2$. Therefore the eigenvalue equation can be written as
\begin{eqnarray}
&&\left[\left(t_2-\frac{\gamma_2}{2}\right)\beta^{-1}+\left(t_1+\frac{\gamma_1}{2}\right)+t_3\beta\right] \nonumber\\
&&\times\left[t_3\beta^{-1}+\left(t_1-\frac{\gamma_1}{2}\right)+\left(t_2+\frac{\gamma_2}{2}\right)\beta\right]=E^2.
\label{eq73a}
\end{eqnarray}
Henceforth we assume $t_2\neq\pm\gamma_2/2$ and $t_3\neq0$, meaning that Eq.~(\ref{eq73a}) is a quartic equation for $\beta$, having four solutions $\beta_j~(j=1,\cdots,4)$. They are numbered so as to satisfy $\left|\beta_1\right|\leq\left|\beta_2\right|\leq\left|\beta_3\right|\leq\left|\beta_4\right|$.
\subsection{\label{sec4-2} Derivation of the boundary equation and the condition for the continuum bands}
\begin{figure}[]
\includegraphics[width=8.5cm]{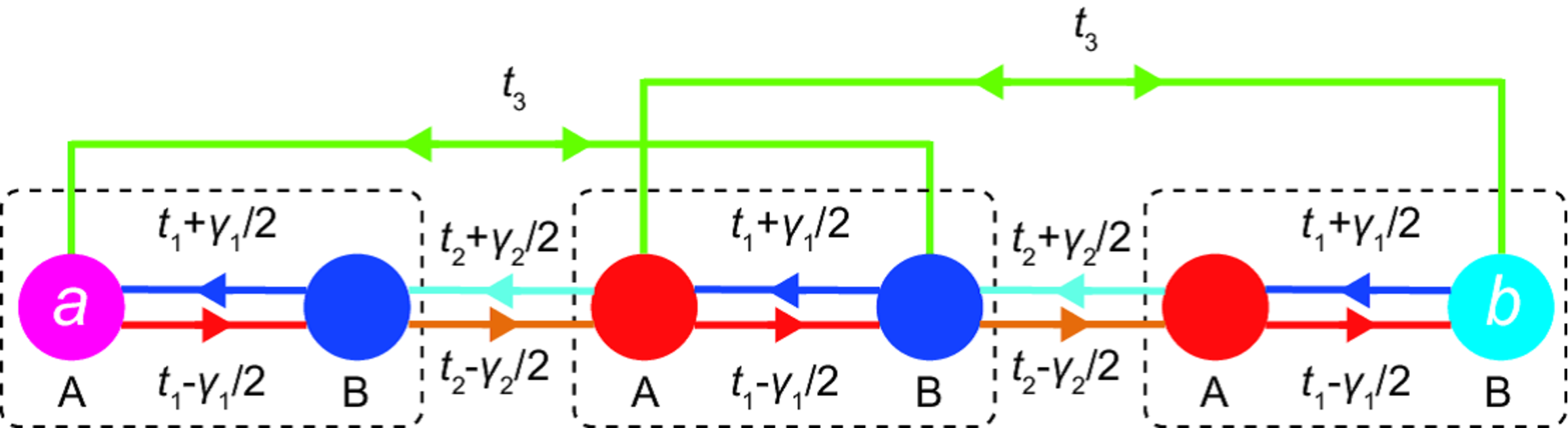}
\caption{\label{fig1b}Schematic figure of the non-Hermitian SSH model with the complex potential $a$ and $b$ on the boundary sites. We show that the number of unit cells is three in this figure.}
\end{figure}
Next we study the non-Hermitian SSH model with $L$ unit cells with the complex potential $a$ and $b$ on the boundary sites as shown in Fig.~\ref{fig1b}. In this subsection, we show that the condition for the continuum bands is indeed given by Eq.~(\ref{eq71a}), independent of $a,b$, and $L$. We write down the boundary condition for $\psi_{n,\mu}$:
\begin{eqnarray}
\left(t_1+\frac{\gamma_1}{2}\right)\psi_{1,{\rm B}}+t_3\psi_{2,{\rm B}}+a\psi_{1,{\rm A}}&=&E\psi_{1,{\rm A}}, \nonumber\\
\left(t_1-\frac{\gamma_1}{2}\right)\psi_{1,{\rm A}}+\left(t_2+\frac{\gamma_2}{2}\right)\psi_{2,{\rm A}}&=&E\psi_{1,{\rm B}}, \nonumber\\
\left(t_2-\frac{\gamma_2}{2}\right)\psi_{L-1,{\rm B}}+\left(t_1+\frac{\gamma_1}{2}\right)\psi_{L,{\rm B}}&=&E\psi_{L,{\rm A}}, \nonumber\\
t_3\psi_{L-1,{\rm A}}+\left(t_1-\frac{\gamma_1}{2}\right)\psi_{L,{\rm A}}+b\psi_{L,{\rm B}}&=&E\psi_{L,{\rm B}}.
\label{eq13b}
\end{eqnarray}
Then, by substituting the general solution
\begin{equation}
\psi_{n,\mu}=\sum_{j=1}^4\left(\beta_j\right)^n\phi_\mu^{\left(j\right)},~\left(\mu={\rm A},{\rm B}\right)
\label{eq23b}
\end{equation}
to Eq.~(\ref{eq13b}), one can obtain four equations for the eight coefficients $\phi_\mu^{\left(j\right)}~\left(j=1,\cdots,4,~\mu={\rm A},{\rm B}\right)$. By recalling that these coefficients satisfy
\begin{eqnarray}
\phi_{\rm A}^{\left(j\right)}&=&\frac{E}{t_3\beta_j^{-1}+\left(t_1-\gamma_1/2\right)+\left(t_2+\gamma_2/2\right)\beta_j}\phi_{\rm B}^{\left(j\right)}, \nonumber\\
\phi_{\rm B}^{\left(j\right)}&=&\frac{E}{\left(t_2-\gamma_2/2\right)\beta_j^{-1}+\left(t_1+\gamma_1/2\right)+t_3\beta_j}\phi_{\rm A}^{\left(j\right)}, \nonumber\\
\label{eq33b}
\end{eqnarray}
from the bulk eigen-equation Eq.~(\ref{eq53a}), we can reduce the problem into four linear equations for the four coefficients $\phi_{\rm A}^{\left(j\right)}~\left(j=1,\cdots,4\right)$. Then the condition for the linear equation to have nontrivial solutions is written as
\begin{eqnarray}
\left| \begin{array}{cccc}
1            & 1            & 1             & 1            \vspace{3pt}\\
f_1          & f_2          & f_3           & f_4          \vspace{3pt}\\
X_1\beta_1^L & X_2\beta_2^L & X_3\beta_3^L  & X_4\beta_4^L \vspace{3pt}\\
g_1\beta_1^L & g_2\beta_2^L & g_3\beta_3^L  & g_4\beta_4^L
\end{array}\right|=0,
\label{eq43b}
\end{eqnarray}
where $X_j$, $f_j$ and $g_j~(j=1,\cdots,4)$ are defined as
\begin{eqnarray}
X_j&=&\frac{\beta_j}{\left(t_2-\gamma_2/2\right)\beta_j^{-1}+\left(t_1+\gamma_1/2\right)+t_3\beta_j}, \nonumber\\
f_j&=&a\beta_j-\left(t_2-\gamma_2/2\right)E\beta_j^{-1}X_j, \nonumber\\
g_j&=&bE\beta_j^{-1}X_j-\left(t_2+\gamma_2/2\right)\beta_j,
\label{eq53b}
\end{eqnarray}
respectively. Furthermore Eq.~(\ref{eq43b}) can be explicitly written as
\begin{eqnarray}
\frac{1}{2}\sum_\sigma{\rm sgn}\left(\sigma\right)\left(f_{\sigma\left(1\right)}-f_{\sigma\left(2\right)}\right)g_{\sigma\left(3\right)}X_{\sigma\left(4\right)}\left(\beta_{\sigma\left(3\right)}\beta_{\sigma\left(4\right)}\right)^L=0, \nonumber\\
\label{eq63b}
\end{eqnarray}
where the sum is taken over all the permutations $\sigma$ for four objects.

Next we demonstrate that the condition for the continuum bands is given by $\left|\beta_2\right|=\left|\beta_3\right|$. We have to consider the condition when the solutions of Eq.~(\ref{eq63b}) are densely distributed for a large $L$. When $\left|\beta_2\right|\neq\left|\beta_3\right|$, the only leading term in Eq.~(\ref{eq63b}) is the term proportional to $\left(\beta_3\beta_4\right)^L$, leading to an equation $\left(f_1-f_2\right)\left(g_3X_4-g_4X_3\right)=0$ in the thermodynamic limit $L\rightarrow\infty$. By combining this equation with the eigenvalue equation (\ref{eq73a}), the eigenenergies are restricted to discrete values, and it cannot represent the continuum bands. On the other hand, when
\begin{equation}
\left|\beta_2\right|=\left|\beta_3\right|
\label{eq73b}
\end{equation}
is satisfied, Eq.~(\ref{eq63b}) has two leading terms, $\left(\beta_2\beta_4\right)^L$ and $\left(\beta_3\beta_4\right)^L$ for a large $L$, and Eq.~(\ref{eq63b}) can be rewritten as
\begin{equation}
\left(\frac{\beta_{2}}{\beta_3}\right)^L=\frac{\left(f_1-f_2\right)\left(g_4X_3-g_3X_4\right)}{\left(f_1-f_3\right)\left(g_4X_2-g_2X_4\right)}.
\label{eq83b}
\end{equation}
For a large $L$, this equation allows a dense set of solutions when the relative phase between $\beta_2$ and $\beta_3$ is continuously changed. Therefore we conclude that Eq.~(\ref{eq73b}) is an appropriate condition for the continuum bands. We here emphasize that Eq.~(\ref{eq73b}) is now independent of any boundary conditions $a$ and $b$, and the system size $L$. If we change the value of the parameters, Eq.~(\ref{eq63b}) changes; nonetheless Eq.~(\ref{eq73b}) remains the same, and together with the eigenvalue equation, we can obtain the continuum bands.

With a special choice of parameters, the degree of the eigenvalue equation may become less than $2M$. For example, when $t_2=-\gamma_2/2$, Eq.~(\ref{eq73a}) becomes a cubic equation for $\beta$. In this case, we can still regard Eq.~(\ref{eq73a}) as a quartic equation by putting $t_2+\gamma_2/2$ to be an infinitesimal. Therefore, by adding another solution $\beta=\infty$ to three solutions for the cubic equation, our result Eq.~(\ref{eq73b}) holds good with four solutions $\beta_1$, $\beta_2$, $\beta_3$, $\beta_4(=\infty)$. Thus, in general, when the degree of the eigenvalue equation is less than $2M$, one can treat it as a limiting case of an algebraic equation of $2M$-th degree by formally adding a solution $\beta=0$ or $\beta=\infty$, depending on the system, and the condition for the continuum bands remains valid.

Another special case is the case of $t_3=0$ in the non-Hermitian SSH model. In this case, the eigenvalue equation becomes a quadratic equation for $\beta$, and it has two solutions. Even in this special case, if the infinitesimally small hopping amplitude $t_3$ is added, the eigenvalue equation has the four solutions $\beta=\beta_1,\beta_2,\beta_3,\beta_4$ satisfying Eq.~(\ref{eq61a}). Then the above condition corresponds to the condition $\beta_1\left(=0\right)\leq\left|\beta_2\right|=\left|\beta_3\right|\leq\beta_4\left(=\infty\right)$ in the limit $t_3\rightarrow0$, meaning that the two solutions $\left(\beta_2,\beta_3\right)$ of the quadratic equation have the same absolute values. This result agrees with the previous work~\cite{Yao2018}.
\subsection{\label{sec4-3}Generalized Brillouin zone}
\begin{figure}[]
\includegraphics[width=8.5cm]{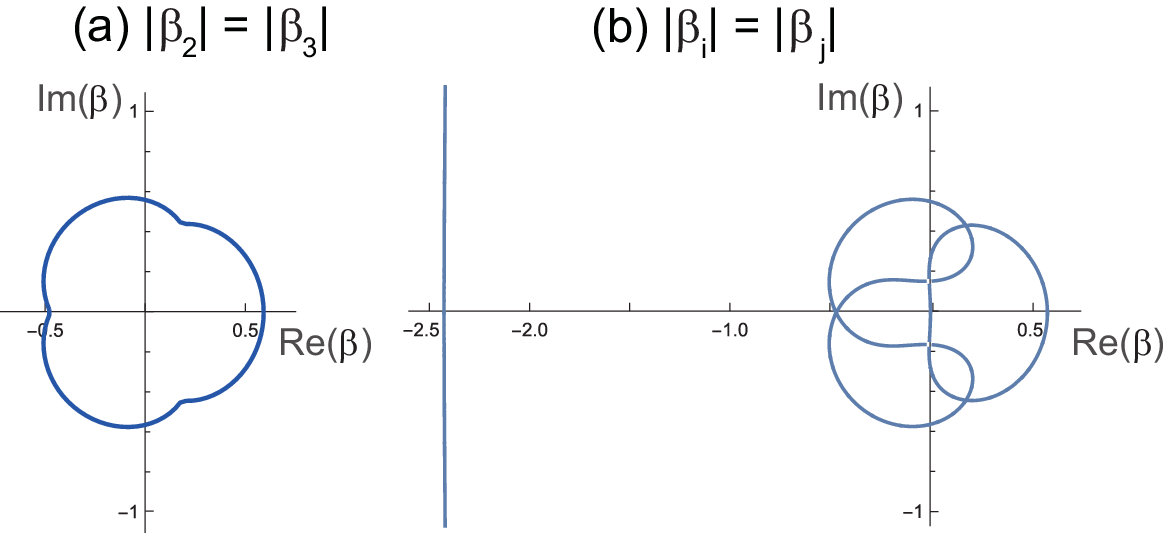}
\caption{\label{fig:BZ-SM}Trajectories of $\beta$. (a) Trajectory of $\beta_2$ and $\beta_3$ with $\left|\beta_2\right|=\left|\beta_3\right|$ and (b) trajectories of $\beta_i$ and $\beta_j$ with $\left|\beta_i\right|=\left|\beta_j\right|~(i\neq j)$. (a) corresponds to the generalized Brillouin zone $C_\beta$. The parameters are same as Fig.~3~(c-1) in the main text: $t_1=0.3,t_2=0.5,t_3=1/5,\gamma_1=5/3$, and $\gamma_2=1/3$.}
\end{figure}

As an example, we show the generalized Brillouin zone $C_\beta$ for parameters $t_1=0.3,t_2=0.5,t_3=1/5,\gamma_1=5/3$, and $\gamma_2=1/3$ in our model. We impose the condition $\left|\beta_2\right|=\left|\beta_3\right|$ and calculate $C_\beta$ as the trajectories of $\beta_2$ and $\beta_3$ shown in Fig.~\ref{fig:BZ-SM}~(a). For comparison, we investigate what happens if we impose a condition $\left|\beta_i\right|=\left|\beta_j\right|$ for some $i$ and $j$ among the four solutions instead. We show the trajectory of $\beta_i$ and $\beta_j$ in Fig.~\ref{fig:BZ-SM}~(b) under the condition $\left|\beta_i\right|=\left|\beta_j\right|$ for some $i$ and $j$. In Ref.~\onlinecite{Yao2018}, it was suggested that $\left|\beta_i\right|=\left|\beta_j\right|$ is a necessary condition for the continuum bands. Nonetheless it is not sufficient to get $C_\beta$, and we should restrict this condition to be $\left|\beta_2\right|=\left|\beta_3\right|$, leading to $C_\beta$ in Fig.~\ref{fig:BZ-SM}~(a).

In some cases, $C_\beta$ can have cusps as seen in Fig.~\ref{fig:BZ-SM}~(a). It appears when three of the four solutions of $\beta$ share the same absolute value. Suppose $\left|\beta_1\right|<\left|\beta_2\right|=\left|\beta_3\right|<\left|\beta_4\right|$, and as we go along $C_\beta$, $\left|\beta_1\right|$ approaches $\left|\beta_2\right|\left(=\left|\beta_3\right|\right)$. Then, when $\left|\beta_1\right|=\left|\beta_2\right|=\left|\beta_3\right|$, the behavior of the solutions satisfying $\left|\beta_2\right|=\left|\beta_3\right|$ changes, and there appears a cusp in $C_\beta$, as one can see from Figs.~\ref{fig:BZ-SM}~(a) and (b).
\subsection{\label{sec4-4}Methods for computing the generalized Brillouin zone}
We explain our method to get the trajectory of $\beta$ satisfying the continuum-band condition (\ref{eq73b}), as shown in Fig.~\ref{fig:BZ-SM}. We first express the eigenvalue equation Eq.~(\ref{eq73a}) as $E^2=f\left(\beta\right)$. Suppose the two solutions $\beta$ and $\beta^\prime$ have the same absolute values: $\left|\beta\right|=\left|\beta^\prime\right|$. Then we have 
\begin{equation}
\beta^\prime=\beta{\rm e}^{i\theta},
\label{eq13d}
\end{equation}
where $\theta$ is real. Then, by taking the difference between two equations:
\begin{equation}
E^2=f\left(\beta\right),~E^2=f\left(\beta{\rm e}^{i\theta}\right),
\label{eq23d}
\end{equation}
we get 
\begin{eqnarray}
&&0=\left(t_2-\frac{\gamma_2}{2}\right)t_3\beta^{-2}\left(1-{\rm e}^{-2i\theta}\right) \nonumber\\
&&+\left[\left(t_1-\frac{\gamma_1}{2}\right)\left(t_2-\frac{\gamma_2}{2}\right)+\left(t_1+\frac{\gamma_1}{2}\right)t_3\right]\beta^{-1}\left(1-{\rm e}^{-i\theta}\right) \nonumber\\
&&+\left[\left(t_1-\frac{\gamma_1}{2}\right)t_3+\left(t_1+\frac{\gamma_1}{2}\right)\left(t_2+\frac{\gamma_2}{2}\right)\right]\beta\left(1-{\rm e}^{i\theta}\right) \nonumber\\
&&+\left(t_2+\frac{\gamma_2}{2}\right)t_3\beta^2\left(1-{\rm e}^{2i\theta}\right).
\label{eq33d}
\end{eqnarray}
This equation allows us to compute $\beta$ for a given value of  $\theta\in\left(0,2\pi\right)$. Then we obtain a set of values of $\beta$ that satisfies $\left|\beta\right|=\left|\beta^\prime\right|$. Here, to obtain the generalized Brillouin zone $C_\beta$, we should further constrain the values of $\beta$ by Eq.~(\ref{eq73b}). Namely the absolute values of $\beta$ and $\beta^\prime$ should be the second and third largest ones among the four solutions. By selecting the values of $\beta$ and $\beta^\prime$ satisfying this condition, we can get $C_\beta$.

%
%

\section{\label{sec5}$Q$ matrix and winding number in 1D non-Hermitian systems with chiral symmetry}
\subsection{\label{sec5-1}Multi-band model}
In this section, we focus on 1D non-Hermitian systems with chiral symmetry which have an arbitrary number of bands, and derive a formula for the winding number. In the following, the systems have a gap around $E=0$, but without exceptional points. Since the systems always have pairs of the eigenenergy $\left(E,-E\right)$ due to chiral symmetry, we can assume that the bands are composed of $N$ occupied bands with $E=-E_i~(i=1,\cdots,N)$ and $N$ unoccupied bands with $E=E_i~(i=1,\cdots,N)$. By taking an appropriate basis, the Hamiltonian in the systems can be written as the block off-diagonal form:
\begin{eqnarray}
{\cal H}\left(\beta\right)=\left( \begin{array}{cc}
0                     & R_+\left(\beta\right) \vspace{5pt}\\
R_-\left(\beta\right) & 0
\end{array}\right),
\label{eq15a}
\end{eqnarray}
where $\beta\equiv{\rm e}^{ik},~k\in{\mathbb C}$ and $R_{\pm}\left(\beta\right)$ are $N\times N$ matrices. The chiral symmetry is expressed as
\begin{equation}
\sigma_z{\cal H}(\beta)=-{\cal H}(\beta)\sigma_z,\ \ 
\sigma_z\equiv\left( \begin{array}{cc}
\bm{1}                     & 0       \vspace{5pt}\\
0                          & -\bm{1}
\end{array}\right),
\label{eq25a}
\end{equation}
where $\bm{1}$ is an $N\times N$ identity matrix. Then the eigenvalue equations for the right and left eigenstates
\begin{eqnarray}
\ket{\psi_{R}\left(\beta\right)}&=&\left( \begin{array}{c}
\ket{a_{R}\left(\beta\right)} \vspace{5pt}\\
\ket{b_{R}\left(\beta\right)}
\end{array}\right), \nonumber \\
\bra{\psi_{L}\left(\beta\right)}&=&\left( \begin{array}{cc}
\bra{a_{L}\left(\beta\right)}, & \bra{a_{R}\left(\beta\right)}
\end{array}\right),
\label{eq35a}
\end{eqnarray}
are given by
\begin{eqnarray}
R_+\left(\beta\right)\ket{b_R\left(\beta\right)}&=&E\ket{a_R\left(\beta\right)}, \nonumber\\
R_-\left(\beta\right)\ket{a_R\left(\beta\right)}&=&E\ket{b_R\left(\beta\right)},
\label{eq45a}
\end{eqnarray}
and
\begin{eqnarray}
\bra{a_L\left(\beta\right)}R_+\left(\beta\right)&=&E\bra{b_L\left(\beta\right)}, \nonumber\\
\bra{b_L\left(\beta\right)}R_-\left(\beta\right)&=&E\bra{a_L\left(\beta\right)},
\label{eq55a}
\end{eqnarray}
respectively. For the right and left eigenstates, one can reduce Eqs.~(\ref{eq45a}) and (\ref{eq55a}) to
\begin{eqnarray}
R_+\left(\beta\right)R_-\left(\beta\right)\ket{a_R\left(\beta\right)}&=&E^2\ket{a_R\left(\beta\right)}, \nonumber\\
\bra{a_L\left(\beta\right)}R_+\left(\beta\right)R_-\left(\beta\right)&=&E^2\bra{a_L\left(\beta\right)},
\label{eq65a}
\end{eqnarray}
respectively. Here we introduce the right and left eigenstates of the $N\times N$ matrix $R_+\left(\beta\right)R_-\left(\beta\right)$ as $\ket{a_{R/L,1}\left(\beta\right)},\cdots,\ket{a_{R/L,N}\left(\beta\right)}$, respectively, and the eigenvalues as $E_1^2\left(\beta\right),\cdots,E_N^2\left(\beta\right)$. Furthermore the right and left eigenstates satisfy
\begin{equation}
\Braket{a_{L,i}\left(\beta\right)|a_{R,j}\left(\beta\right)}=\delta_{ij}
\label{eq75a}
\end{equation}
since one can take the biorthogonal basis~\cite{Brody2014,Shen2018}. Therefore we can obtain the biorthogonal eigenstates of the Hamiltonian Eq.~(\ref{eq15a}) in the occupied bands for $i=1,\cdots,N$ as
\begin{eqnarray}
\ket{\psi_{R,i}\left(\beta\right)}&=&\frac{1}{\sqrt{2}}\left( \begin{array}{c}
\ket{a_{R,i}\left(\beta\right)} \vspace{5pt}\\
-\displaystyle\frac{1}{E_i\left(\beta\right)}R_-\left(\beta\right)\ket{a_{R,i}\left(\beta\right)}
\end{array}\right), \nonumber\\
\bra{\psi_{L,i}\left(\beta\right)}&=&\frac{1}{\sqrt{2}}\left( \begin{array}{cc}
\bra{a_{L,i}\left(\beta\right)}, & -\displaystyle\frac{1}{E_i\left(\beta\right)}\bra{a_{L,i}}R_+\left(\beta\right)
\end{array}\right). \nonumber\\
\label{eq85a}
\end{eqnarray}
Then the unoccupied eigenstates are given by $\ket{\widetilde{\psi}_{R/L,i}\left(\beta\right)}\equiv\sigma_z\ket{\psi_{R/L,i}\left(\beta\right)}$. In the systems, the $Q$ matrix $Q\left(\beta\right)$ can be defined as 
\begin{equation}
Q\left(\beta\right)=\sum_{i=1}^N\left(\ket{\widetilde{\psi}_{R,i}\left(\beta\right)}\bra{\widetilde{\psi}_{L,i}\left(\beta\right)}-\ket{\psi_{R,i}\left(\beta\right)}\bra{\psi_{L,i}\left(\beta\right)}\right)
\label{eq95a}
\end{equation}
and in the matrix form, it can be written as
\begin{widetext}
\begin{eqnarray}
Q\left(\beta\right)=\sum_{i=1}^N\frac{1}{E_i\left(\beta\right)}\left( \begin{array}{cc}
O                                                                                   & \ket{a_{R,i}\left(\beta\right)}\bra{a_{L,i}\left(\beta\right)}R_+\left(\beta\right) \vspace{5pt}\\
R_-\left(\beta\right)\ket{a_{R,i}\left(\beta\right)}\bra{a_{L,i}\left(\beta\right)} & O
\end{array}\right).
\label{eq105a}
\end{eqnarray}
\end{widetext}
Here we define the matrix $q\left(\beta\right)$ and $q^{-1}\left(\beta\right)$ as
\begin{eqnarray}
q\left(\beta\right)&\equiv&\sum_{i=1}^N\frac{1}{E_i\left(\beta\right)}\ket{a_{R,i}\left(\beta\right)}\bra{a_{L,i}\left(\beta\right)}R_+\left(\beta\right), \nonumber\\
q^{-1}\left(\beta\right)&=&\sum_{i=1}^N\frac{1}{E_i\left(\beta\right)}R_-\left(\beta\right)\ket{a_{R,i}\left(\beta\right)}\bra{a_{L,i}\left(\beta\right)}, \nonumber\\
\label{eq115a}
\end{eqnarray}
respectively. As a result, one can get the winding number $w$~\cite{Ryu2010,Yao2018} as
\begin{equation}
w=\frac{i}{2\pi}\int_{C_\beta}{\rm Tr}\left[{\rm d} q~q^{-1}\left(\beta\right)\right].
\label{eq125a}
\end{equation}
\subsection{\label{sec5-2}Two-band model}
In particular, we can explicitly write Eq.~(\ref{eq125a}) in a two-band model. The Hamiltonian can be written as
\begin{equation}
{\cal H}\left(\beta\right)=R_+\left(\beta\right)\sigma_++R_-\left(\beta\right)\sigma_-,
\label{eq15b}
\end{equation}
where $\sigma_\pm=\left(\sigma_x\pm i\sigma_y\right)/2$. Then the eigenvalues are given by
\begin{equation}
E_\pm\left(\beta\right)=\pm\sqrt{R_+\left(\beta\right)R_-\left(\beta\right)}.
\label{eq25b}
\end{equation}
The right and left eigenstates can be written down as
\begin{eqnarray}
\ket{u_{R,+}}&=&\frac{1}{\sqrt{2}\sqrt{R_+R_-}}\left( \begin{array}{c}
R_+ \vspace{5pt}\\
\sqrt{R_+R_-}
\end{array}\right), \nonumber\\
\ket{u_{R,-}}&=&\frac{1}{\sqrt{2}\sqrt{R_+R_-}}\left( \begin{array}{c}
R_+ \vspace{5pt}\\
-\sqrt{R_+R_-}
\end{array}\right), \nonumber\\
\bra{u_{L,+}}&=&\frac{1}{\sqrt{2}\sqrt{R_+R_-}}\left( \begin{array}{cc}
R_-, & \sqrt{R_+R_-}
\end{array}\right), \nonumber\\
\bra{u_{L,-}}&=&\frac{1}{\sqrt{2}\sqrt{R_+R_-}}\left( \begin{array}{cc}
R_-, & -\sqrt{R_+R_-}
\end{array}\right),
\label{eq35b}
\end{eqnarray}
respectively. The subscript $+/-$ means that the eigenstates with $+/-$ have the eigenvalues $E_+$ or $E_-$, respectively. From Eq.~(\ref{eq35b}), the $Q$ matrix $Q\left(\beta\right)$ is given by
\begin{eqnarray}
Q\left(\beta\right)&=&\ket{u_{R,+}\left(\beta\right)}\bra{u_{L,+}\left(\beta\right)}-\ket{u_{R,-}\left(\beta\right)}\bra{u_{L,-}\left(\beta\right)} \nonumber\\
&=&\frac{1}{\sqrt{R_+\left(\beta\right)R_-\left(\beta\right)}}\left( \begin{array}{cc}
0                     & R_+\left(\beta\right) \\
R_-\left(\beta\right) & 0
\end{array}\right).
\label{eq45b}
\end{eqnarray}
Therefore we obtain $q=R_+/\sqrt{R_+R_-}$, and the winding number $w$ as
\begin{eqnarray}
w&=&\frac{i}{2\pi}\int_{C_\beta}{\rm d}q~q^{-1}\left(\beta\right) \nonumber\\
&=&\frac{i}{2\pi}\int_{C_\beta}{\rm d}\log q\left(\beta\right) \nonumber\\
&=&-\frac{1}{2\pi}\left[\arg q\left(\beta\right)\right]_{C_\beta} \nonumber\\
&=&-\frac{1}{2\pi}\frac{\left[\arg R_+\left(\beta\right)-\arg R_-\left(\beta\right)\right]_{C_\beta}}{2}.
\label{eq55b}
\end{eqnarray}

Thus the winding number $w$ is determined by the change of the phase of $R_\pm\left(\beta\right)$ when $\beta$ goes along the generalized Brillouin zone $C_\beta$. On a complex plane, let $\ell_{\pm}$ denote the loops drawn by $R_{\pm}\left(\beta\right)$ when $\beta$ goes along $C_\beta$ in the counterclockwise way. Then $w$ is determined by the number of times that $\ell_{\pm}$ surround the origin ${\cal O}$. When neither $\ell_+$ nor $\ell_-$ surrounds ${\cal O}$, $w$ is zero. It takes a non-zero value when they simultaneously surround ${\cal O}$.

Here we can show the bulk-edge correspondence in this case. If parameters of the system can be continuously changed without closing the gap, the winding number $w$ does not change, and the topology of the systems remains invariant. Namely, if the systems after changing the values of the parameters have the zero-energy edge states, one can conclude that the original systems also have the zero-energy edge states. This can be proved even in non-Hermitian cases, following the proof in Hermitian systems~\cite{Ryu2002}.

Suppose we change the parameters of the system continuously, and at some values of the parameters, $\ell_+$ and $\ell_-$ simultaneously pass the origin ${\cal O}$ on the ${\bm R}$ plane. At this time, the gap closing occurs because the energy eigenvalues are given by Eq.~(\ref{eq25b}).

In Hermitian systems, $R_+^\ast=R_-$ holds, and two loops $\ell_+$ and $\ell_-$ are related by complex conjugation, so 
\begin{equation}
\left[\arg R_-\left(\beta\right)\right]_{C_\beta}=-\left[\arg R_+\left(\beta\right)\right]_{C_\beta}.
\label{eq65b}
\end{equation}
In contrast with these cases, in some non-Hermitian systems, only one of the two loops $\ell_+$ and $\ell_-$ passes the origin ${\cal O}$. In that case, only one of the two values $R_+$ or $R_-$ become $0$, leading to $E=0$, and the Hamiltonian is written as the Jordan normal form. It represents that the system has exceptional points. In Figs.~3~(a) and 4~(a) in the main text, the phase with the exceptional point extends over a finite region due to the following reason. When $R_+=0$ (or $R_-=0$) has non-real solutions, these two solutions $\beta_1^+$ and $\beta_2^+$ (or $\beta_1^-$ and $\beta_2^-$) should be complex conjugate, having the same absolute value, because all the coefficients of $R_+=0$ and $R_-=0$ are real. This persists as long as the following condition are satisfied: $R_+=0$ (or $R_-=0$) has non-real two solutions satisfying $\left|\beta_1^-\right|<\left|\beta_1^+\right|=\left|\beta_2^+\right|<\left|\beta_2^-\right|$ (or $\left|\beta_1^+\right|<\left|\beta_1^-\right|=\left|\beta_2^-\right|<\left|\beta_2^+\right|$). We note that the winding number is not well-defined when the system has the exceptional pints because the gap is closed.

%
%

\section{\label{sec6}Calculation on another model}
\begin{figure}[]
\includegraphics[width=8.5cm]{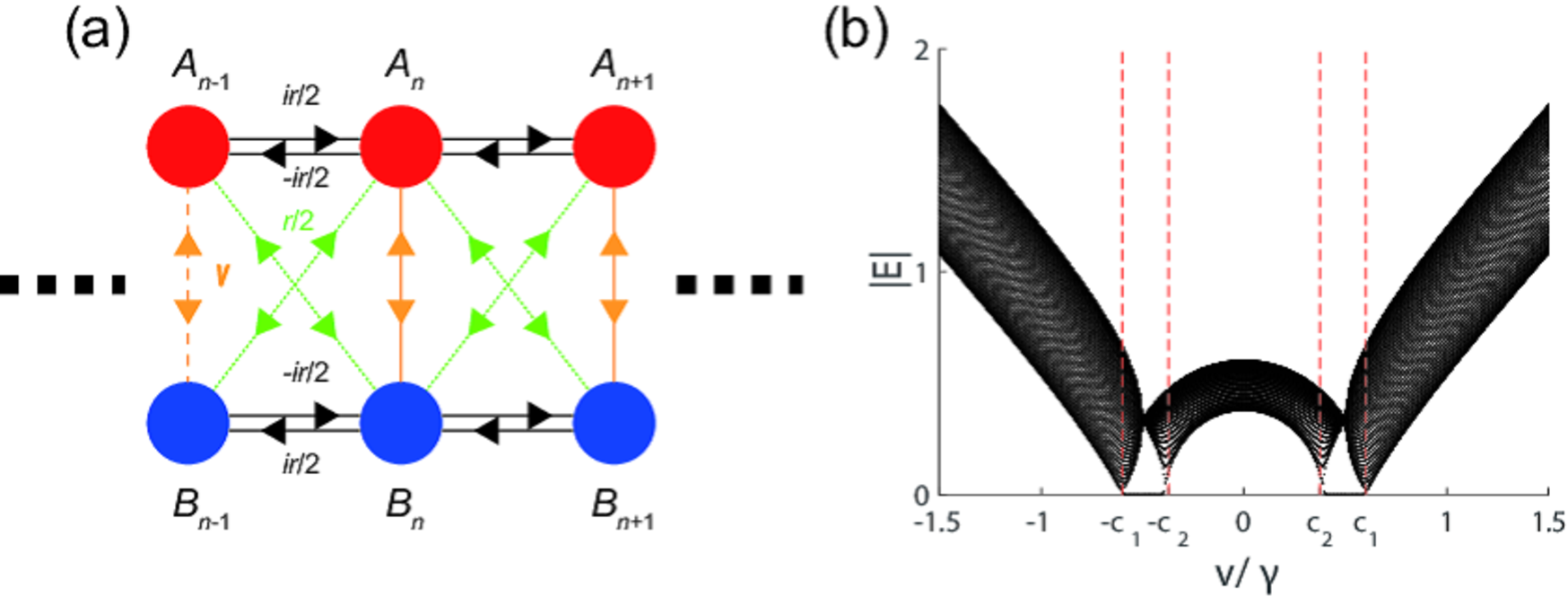}
\caption{\label{fig2}(a) Schematic figure of the tight-binding model proposed in the previous work~\cite{Lee2016}. This model includes the gain on sublattice A and the loss on sublattice B. (b) Energy bands in a finite open chain of the model given by Eq.~(\ref{eq16}). We set the parameter as $r/\gamma=1/3$ and the number of unit cells as $L=50$. The gap closes at $v/\gamma=\pm c_1=\pm\sqrt{13}/6$ and $v/\gamma=\pm c_2=\pm\sqrt{5}/6$. The red dashed lines represent the gap-closing points in the continuum bands.}
\end{figure}
In this section, we investigate the 1D non-Hermitian tight-binding model proposed in Ref.~\onlinecite{Lee2016} as shown in Fig.~\ref{fig2} (a) in terms of the complex Bloch wave number $k\in{\mathbb C}$.

First of all, we can write the Hamiltonian in this model as
\begin{eqnarray}
H&=&\sum_n\left[v\left(c_{n,{\rm A}}^\dag c_{n,{\rm B}}+c_{n,{\rm B}}^\dag c_{n,{\rm A}}\right)\right. \nonumber\\
&+&\frac{ir}{2}\left(c_{n+1,{\rm A}}^\dag c_{n,{\rm A}}-c_{n,{\rm A}}^\dag c_{n+1,{\rm A}}-c_{n+1,{\rm B}}^\dag c_{n,{\rm B}}+c_{n,{\rm B}}^\dag c_{n+1,{\rm B}}\right) \nonumber\\
&+&\frac{r}{2}\left(c_{n+1,{\rm A}}^\dag c_{n,{\rm B}}+c_{n,{\rm B}}^\dag c_{n+1,{\rm A}}+c_{n+1,{\rm B}}^\dag c_{n,{\rm A}}+c_{n,{\rm A}}^\dag c_{n+1,{\rm B}}\right) \nonumber\\
&+&\left.\frac{i\gamma}{2}\left(c_{n,{\rm A}}^\dag c_{n,{\rm A}}-c_{n,{\rm B}}^\dag c_{n,{\rm B}}\right)\right],
\label{eq16}
\end{eqnarray}
where we take $r$, $v$, and $\gamma$ to be real, and henceforth, we set $\left|r/\gamma\right|\leq1/2$. When the eigenvector of the real-space eigen-equation is written as $\ket{\psi}=\left(\cdots,\psi_{1,{\rm A}},\psi_{1,{\rm B}},\psi_{2,{\rm A}},\psi_{2,{\rm B}},\cdots\right)^{\rm T}$, the equation for $\psi_{n,\mu}$ can be written as
\begin{eqnarray}
&&\frac{ir}{2}\left(\psi_{n-1,{\rm A}}-\psi_{n+1,{\rm A}}\right)+\frac{i\gamma}{2}\psi_{n,{\rm A}} \nonumber\\
&&+\frac{r}{2}\left(\psi_{n-1,{\rm B}}+\psi_{n+1,{\rm B}}\right)+v\psi_{n,{\rm B}}=E\psi_{n,{\rm A}}, \nonumber\\
&&-\frac{ir}{2}\left(\psi_{n-1,{\rm B}}-\psi_{n+1,{\rm B}}\right)-\frac{i\gamma}{2}\psi_{n,{\rm B}} \nonumber\\
&&+\frac{r}{2}\left(\psi_{n-1,{\rm A}}+\psi_{n+1,{\rm A}}\right)+v\psi_{n,{\rm A}}=E\psi_{n,{\rm B}}. 
\label{eq26}
\end{eqnarray}
Here we take an ansatz as a linear combination:
\begin{equation}
\psi_{n,\mu}=\sum_j\phi_{n,\mu}^{\left(j\right)},~(\mu={\rm A},{\rm B}),
\label{eq36}
\end{equation}
where $\phi_{n,{\rm A}}^{\left(j\right)}$ and $\phi_{n,{\rm B}}^{\left(j\right)}$ take the exponential form
\begin{equation}
\left(\phi^{\left(j\right)}_{n,{\rm A}},\phi^{\left(j\right)}_{n,{\rm B}}\right)=\left(\beta_j\right)^n\left(\phi^{\left(j\right)}_{\rm A},\phi^{\left(j\right)}_{\rm B}\right).
\label{eq46}
\end{equation}
By substituting the exponential form Eq.~(\ref{eq46}) into Eq.~(\ref{eq26}), one can obtain
\begin{eqnarray}
\left[-\frac{ir}{2}\left(\beta-\beta^{-1}\right)+\frac{i\gamma}{2}\right]\phi_{\rm A}+\left[\frac{r}{2}\left(\beta+\beta^{-1}\right)+v\right]\phi_{\rm B}&=&E\phi_{\rm A}, \nonumber\\
\left[\frac{ir}{2}\left(\beta-\beta^{-1}\right)-\frac{i\gamma}{2}\right]\phi_{\rm B}+\left[\frac{r}{2}\left(\beta+\beta^{-1}\right)+v\right]\phi_{\rm A}&=&E\phi_{\rm B}, \nonumber\\
\label{eq56}
\end{eqnarray}
and the generalized Bloch Hamiltonian ${\cal H}\left(\beta\right)$:
\begin{eqnarray}
{\cal H}\left(\beta\right)=\left[v+\frac{r}{2}\left(\beta+\beta^{-1}\right)\right]\sigma_x+\left[-\frac{ir}{2}\left(\beta-\beta^{-1}\right)+\frac{i\gamma}{2}\right]\sigma_z. \nonumber\\
\label{eq66}
\end{eqnarray}
Therefore the eigenvalue equation is given by
\begin{eqnarray}
r\left(v+\frac{\gamma}{2}\right)\beta+\left(r^2+v^2-\frac{\gamma^2}{4}-E^2\right)+r\left(v-\frac{\gamma}{2}\right)\beta^{-1}=0. \nonumber\\
\label{eq76}
\end{eqnarray}
The condition for the continuum bands can be written as $\left|\beta_1\right|=\left|\beta_2\right|$ since Eq.~(\ref{eq76}) is a quadratic equation for $\beta$. In this case, the generalized Brillouin zone becomes a circle with the radius
\begin{equation}
R\equiv\left|\beta_{1,2}\right|=\sqrt{\left|\frac{v-\gamma/2}{v+\gamma/2}\right|}
\label{eq86}
\end{equation}
which is given by Vieta's formula~\cite{Yao2018}. By substituting $\beta=R{\rm e}^{i\varphi},~\varphi\in{\mathbb R}$ to Eq.~(\ref{eq76}), we can get the gap-closing point as
\begin{equation}
\frac{v}{\gamma}=\pm\sqrt{\frac{1}{4}\pm\left(\frac{r}{\gamma}\right)^2}.
\label{eq96}
\end{equation}
Therefore the zero-energy edge states are expected to appear in the parameter region $v/\gamma\in\left[-\sqrt{1/4+\left(r/\gamma\right)^2},-\sqrt{1/4-\left(r/\gamma\right)^2}\right]$ and $v/\gamma\in\left[\sqrt{1/4-\left(r/\gamma\right)^2},\sqrt{1/4+\left(r/\gamma\right)^2}\right]$ because the winding number is shown to be unity there, as a consequence of the previous discussion. Indeed we confirm the appearance of these zero-energy edge states in a finite open chain with $r/\gamma=1/3$ within these region. The gap closes at $v/\gamma=\pm\sqrt{5}/6,\pm\sqrt{13}/6$ in the continuum bands, and the zero-energy edge states appear in $v/\gamma\in\left[-\sqrt{13}/6,-\sqrt{5}/6\right]$ and $v/\gamma\in\left[\sqrt{5}/6,\sqrt{13}/6\right]$, which perfectly agrees with our conclusion of energy levels in a finite open chain (Fig.~\ref{fig2}~(b)). Thus, by our non-Bloch band theory, the bulk-edge correspondence is fully shown in the same sense as in the Hermitian systems. Namely the closing of the gap in the open chain is explained by Eq.~(\ref{eq96}). At these points, the winding number defined in our theory changes between zero and unity, which exactly coincides with absence and presence of zero-energy in-gap states. These aspects have not been understood within theories with the real Bloch wave number~\cite{Lee2016}.

The tight-binding model in Fig.~\ref{fig2}~(a) can be regarded as the non-Hermitian SSH model~\cite{Yao2018}. Indeed, by the unitary transformation
\begin{equation}
\sigma_x\rightarrow\sigma_x,~\sigma_y\rightarrow-\sigma_z,~\sigma_z\rightarrow\sigma_y,
\label{eq106}
\end{equation}
the generalized Bloch Hamiltonian Eq.~(\ref{eq66}) can be rewritten as
\begin{equation}
{\cal H}\left(\beta\right)\rightarrow\left(v+\frac{\gamma}{2}+r\beta^{-1}\right)\sigma_++\left(v-\frac{\gamma}{2}+r\beta\right)\sigma_-,
\label{eq116}
\end{equation}
where $\sigma_\pm=\left(\sigma_x\pm i\sigma_y\right)/2$. We note that the asymmetric hopping amplitude is caused by the gain and loss in the original system. This Hamiltonian is reduced to the non-Hermitian SSH model as shown in Fig.~\ref{fig1a} with $t_3=\gamma_2=0$.
%

\providecommand{\noopsort}[1]{}\providecommand{\singleletter}[1]{#1}%